\documentclass[sigconf,authorversion]{acmart}

\usepackage{multirow}
\usepackage{subfigure}

\AtBeginDocument{%
  \providecommand\BibTeX{{%
    \normalfont B\kern-0.5em{\scshape i\kern-0.25em b}\kern-0.8em\TeX}}}

\copyrightyear{2021}
\acmYear{2021}
\setcopyright{acmcopyright}\acmConference[CHI '21]{CHI Conference on Human Factors in Computing Systems}{May 8--13, 2021}{Yokohama, Japan}
\acmBooktitle{CHI Conference on Human Factors in Computing Systems (CHI '21), May 8--13, 2021, Yokohama, Japan}
\acmPrice{15.00}
\acmDOI{10.1145/3411764.3445596}
\acmISBN{978-1-4503-8096-6/21/05}

\begin{document}

\title{Digital Transformations of Classrooms in Virtual Reality}

\author{Hong Gao}
\authornote{Both authors contributed equally to this research.}
\email{hong.gao@informatik.uni-tuebingen.de}
\author{Efe Bozkir}
\authornotemark[1]
\email{efe.bozkir@uni-tuebingen.de}
\affiliation{%
  \department{Human-Computer Interaction}
  \institution{University of T{\"u}bingen}
  \city{T{\"u}bingen}
  \country{Germany}
}

\author{Lisa Hasenbein}
\affiliation{%
  \department{Hector Research Institute of Education Sciences and Psychology}
  \institution{University of T{\"u}bingen}
  \city{T{\"u}bingen}
  \country{Germany}}
\email{lisa.hasenbein@uni-tuebingen.de}

\author{Jens-Uwe Hahn}
\affiliation{%
  \institution{Hochschule der Medien Stuttgart}
  \city{Stuttgart}
  \country{Germany}}
\email{hahn@hdm-stuttgart.de}

\author{Richard G{\"o}llner}
\affiliation{%
 \department{Hector Research Institute of Education Sciences and Psychology}
 \institution{University of T{\"u}bingen}
  \city{T{\"u}bingen}
  \country{Germany}}
\email{richard.goellner@uni-tuebingen.de}

\author{Enkelejda Kasneci}
\affiliation{%
 \department{Human-Computer Interaction}
 \institution{University of T{\"u}bingen}
  \city{T{\"u}bingen}
  \country{Germany}}
\email{enkelejda.kasneci@uni-tuebingen.de}

\renewcommand{\shortauthors}{Gao and Bozkir, et al.}

\begin{abstract}
With rapid developments in consumer-level head-mounted displays and computer graphics, immersive VR has the potential to take online and remote learning closer to real-world settings. However, the effects of such digital transformations on learners, particularly for VR, have not been evaluated in depth. This work investigates the interaction-related effects of sitting positions of learners, visualization styles of peer-learners and teachers, and hand-raising behaviors of virtual peer-learners on learners in an immersive VR classroom, using eye tracking data. Our results indicate that learners sitting in the back of the virtual classroom may have difficulties extracting information. Additionally, we find indications that learners engage with lectures more efficiently if virtual avatars are visualized with realistic styles. Lastly, we find different eye movement behaviors towards different performance levels of virtual peer-learners, which should be investigated further. Our findings present an important baseline for design decisions for VR classrooms.
\end{abstract}



\begin{teaserfigure}
  \includegraphics[width=\textwidth]{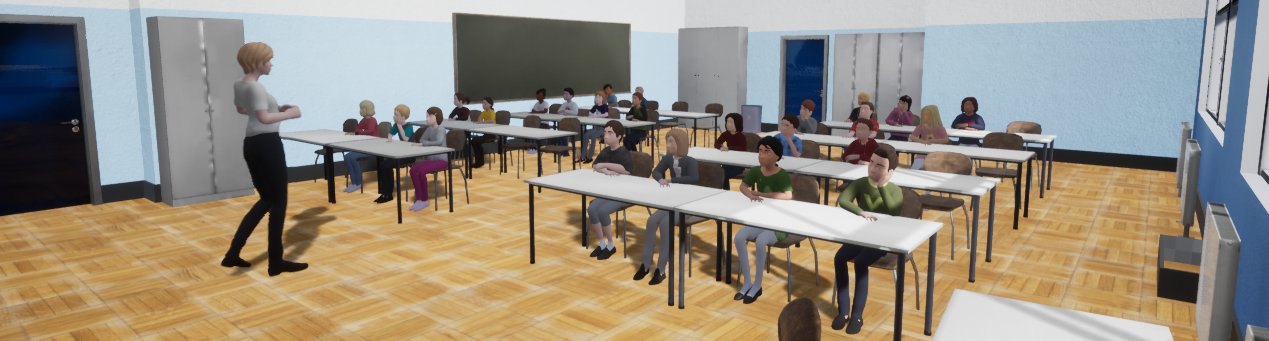}
  \caption{Immersive virtual reality classroom.}
  \Description{Virtual classroom including virtual peer-learners and teacher during the lecture. The teacher stands in the classroom, whereas virtual peer-learners sit and attend the lecture.}
  \label{fig:teaser}
\end{teaserfigure}

\maketitle

\section{Introduction}
Recently, many universities and schools have switched to online teaching due to the COVID-19 pandemic. Online and remote learning may become more prevalent in the near future. However, one of the disadvantages of teaching and learning in such ways compared to conventional classroom-based settings is the limited social interaction with teachers and peer-learners. As this may demotivate learners in the long term, better social engagement providing solutions such as immersive virtual reality (IVR) can be used for teaching and learning. Next-generation VR platforms such as Engage\footnote{https://engagevr.io/} or Mozilla Hubs\footnote{https://hubs.mozilla.com/} may offer better social engagement for learners in the virtual environments; however, the effects of such environments on learners have to be better investigated. In addition to the opportunity to provide more efficient social engagement configurations, VR also enables building and evaluating situations that are difficult to set up in real life (e.g., due to the privacy-related concerns or current availability).

While VR technology has a long history in the education domain~\cite{vreducationdomain, vrandeducation}, the current availability of consumer-grade head-mounted displays (HMDs) allows for the creation of immersive experiences at a reasonable cost, making it possible to employ immersive personalized VR experiences in classrooms in the near future~\cite{riftart}. However, the digital transformations of classrooms reflect an important and critical step when developing VR environments for learning purposes and require further research. A unique opportunity to understand the gaze-based behavior, and consequently, attention distribution of learners in such VR settings is provided through the analysis of the eye movement of learners~\cite{vr_eyetracking_review}. Since some of the high-end HMDs already consist of integrated eye trackers, it does not require extensive effort to extract eye movement patterns during simulations in VR. A thorough analysis of the eye movements allows to infer information on the users going beyond the gaze position, for example stress~\cite{stress_vr}, cognitive load~\cite{bozkir2019person}, visual attention~\cite{Bozkir}, evaluation and diagnosis of diseases~\cite{7829437}, future gaze locations~\cite{8998375}, or training evaluation~\cite{8448290}. In the virtual classroom, this rich source of information could even be combined with the virtual teachers' attention, similar to real-world classrooms~\cite{Sumer_2018_CVPR_Workshops,goldberg2019attentive}, to design more responsive and engaging learning environments.

In this study, we design an immersive VR classroom that is similar to a real classroom, enabling students to perceive an immersive virtual classroom experience. We focus on exploring the impact of the digital transformation from the classroom to immersive VR on learners by analyzing their eye movements. For this purpose, three design factors are studied, including sitting positions of the participating students, different visualization styles of the virtual peer-learners and teachers, and different performance levels of virtual peer-learners with different hand-raising behaviors. Figure ~\ref{fig:teaser} shows the overall design of the virtual classroom. Consequently, our main contributions are as follows.

\begin{itemize}
\item We design an immersive VR classroom and conduct a user study to enable students to virtually perceive ``interactive'' learning.
\item We analyze the effect of different sitting positions on learners, including sitting in the front and back. We find significantly different effects in fixation and saccade durations, and saccade amplitudes in relation to the sitting position.
\item We evaluate the effect of different visualization styles of virtual avatars on learners including cartoon and realistic styles and find significantly different effects in fixation and saccade durations, and pupil diameters.
\item We assess the effect of different performance levels of virtual peer-learners on learners by evaluating various hand-raising percentages, and find significant effects particularly in pupil diameters and number of eye fixations.
\end{itemize}

\section{Related Work}
\label{sec:related_work}
As head-mounted displays (HMDs) and related hardware become more accessible and affordable, VR technology may become an important factor in the educational domain, particularly given its provided immersion and potential for teaching~\cite{review_ivr_education,youngblut1998educational}. Various recent works on VR and education indicate that VR may offer significant advantages for learning and teaching. For instance, based on the post-session knowledge tests, both augmented and virtual reality (AR/VR) are found to promote intrinsic benefits such as increasing learners' immersion and engagement when used for learning structural anatomy~\cite{doi:10.1002/ase.1696}. In~\cite{alhalabi2016virtual}, the impact of VR systems on student achievements in engineering colleges was investigated by evaluating the results of post-quizzes and the results show that VR conditions present significant advantages when compared to no-VR conditions since students improve their performance, which indicates that VR can successfully support teaching engineering classes. Additionally, VR was also evaluated to help teachers develop specific skills that can be helpful in their teaching processes~\cite{lambvirtual}. In addition to teaching and learning processes, another aspect under evaluation concerns the types of virtual environment configurations that are used not only for learning, but also for exploring immersion, motivation, and interaction. To this end, different types of VR setups have been studied. ~\cite{riftart} introduced an immersive VR tool to support teaching and studying art history, which indicates, when used for high-school students, an increased motivation towards art history. ~\cite{smartphone_vr} explored the possibility of using low-cost VR setups to improve daily classroom teaching by using a smartphone-based VR system. According to the evaluations using pre- and post tests, the proposed VR setup helps students perform better compared to traditional teaching using whiteboard and slides. Furthermore, HMD-based VR environment was studied in an elementary classroom for teachers to guide their students in exploring learning elements in immersive virtual field trips~\cite{207ed}. It has been concluded that students' motivation was enhanced after the virtual field trips. Overall, such works imply that while increasing motivation and engagement, different types of VR environments provide plenty of benefits and can be used to assist learning and teaching processes by providing users with immersive experiences.

One disadvantage of such VR and online learning tools is that learners' motivation and performance may be affected by lack of social interaction~\cite{mooc_social_interaction}, peer accompaniment~\cite{doi:10.1177/1052562904271199}, or immersion~\cite{motivation_immersion}. Furthermore, realism in immersive environments can have various implications~\cite{realism}, related to both learning and interaction. To address these issues, several works have focused on how to provide more realistic and immersive environments. For example, ~\cite{vrclassroomconstructivist} discusses the design of the VR environments for classrooms by replicating real learning conditions and enhancing learning through real-time interaction between learners and instructors. Furthermore, ~\cite{8797708} constructed virtual classmates by synthesizing previous learners' time-anchored comments and indicates that when students are accompanied by a small number of virtual peer-learners built with prior learners' comments, their learning outcomes are improved. In addition to virtual peer-learners, the presence of virtual instructors may also have an impact on learning in VR. ~\cite{livehumanrole} investigated this and reports that learners engaged more with the environment and progressed further with the interaction prompts when a virtual instructor was provided. These works and findings indicate that the styles and types of virtual agents in the virtual environments may have several effects on students' attention and perception during immersion and should be taken into account. The evaluation of real-time visual attention towards similar configurations, which could be carried out using sensors such as eye trackers, may not only help to understand learning processes but also provide empirical insights about interactions during virtual classes for digital transformations of classrooms in VR.

From immersion and interaction point of view, video teleconferencing systems share similar goals with the VR classrooms as such systems enable people to experience highly immersive and interactive environments~\cite{teleconference_VR} and have been studied in the VR context as well. For example, ~\cite{teleconference_immersion} proposed a video teleconference experience using a VR headset and found that the sense of immersion and feeling of presence of a remote person increases with VR. Furthermore, different mixed reality (MR)-based 3D collaborative mediums were studied in terms of teleconference backgrounds and user visualization styles~\cite{teleconference_twoaspects}. The real background scene and realistically constructed avatars promote a higher sense of co-presence. Low-cost setups were investigated also for real-time VR teleconferencing~\cite{teleconference_realtime}, as it was done for VR learning environments and it is found that it is possible to improve image quality using headsets in these setups. The possibility of having low-cost setups may become an important factor in the future when accessibility and extensive usage of everyday VR environments for learning~\cite{alhalabi2016virtual} and interaction~\cite{vrclassroomconstructivist} are considered.

In general, while the visualization styles and rendering are considered to affect learners' perception and attention, in virtual learning environments particularly in IVR classrooms, other design factors are also important for attention-related tasks. For instance, ~\cite{bailenson_et_al_2008} has studied the effect of being closer to the teacher, being in the teacher's field of view (FOV), and the availability of virtual co-learners in virtual classrooms. In particular, the authors found that students learn more if they are closer to the teacher and by being in the center of the teacher's FOV. In addition, when no co-learners or co-learners who have positive attitudes towards the lecture (e.g., looking at the teacher or taking notes) are available, students learn more information about the lecture instead of the virtual room. Gazing time was approximated according to the time students kept the virtual teacher in their FOVs; however, real-time gaze information was missing during the experiments. Exact gazing patterns and different eye movement events during learning are particularly needed for understanding moment-to-moment visual behaviors of students. In another work, ~\cite{blume_et_al_18} studied the effect of the sitting position on attention-deficit/hyperactivity disorder (ADHD) experiencing students in such classrooms and found indications that front-seated students are affected positively by this configuration in terms of learning. However, similar to ~\cite{bailenson_et_al_2008}, the authors did not have gaze information available but identified that the evaluation of eye movements may provide additional insights during learning, particularly in terms of real-time visual interaction, when learning and cognitive processes are taken into consideration. In addition, eye movements are also considered as choice of measurements to study visual perception during learning~\cite{holmqvist_book_eye_tracking,Jarodzka_Holmqvist_Gruber_2017}. ~\cite{DazOrueta2014AULAVR} and ~\cite{Seo2019JointAV} have studied attention measures and social interaction in similar setups using continuous performance tests and head movements, respectively. The latter work has used head movements as a proxy for visual attention and found that head movements shift between target and interaction partner. This finding partly supports the finding of ~\cite{livehumanrole} that the learners' engagement increases when a virtual instructor is presented. However, both works lack eye movement measurements. As also reported by ~\cite{Seo2019JointAV}, eye movements should be examined along with head movements to understand attention and interaction more in-depth, since eyes can move differently. In addition, ~\cite{Nolin2016ClinicaVRCA} studied the relationship between performance, sense of presence, and cybersickness, whereas ~\cite{mangalmurti_2020} examined attention, more particularly ADHD with continuous performance task in a virtual classroom. However, both works are more in the clinical domain, which are relatively different from an everyday classroom setup. ~\cite{rizzo_bowerly_buckwalter_klimchuk_mitura_parsons_2009} provides a general overview more from clinical perspective. Lastly, although has not been studied extensively in VR yet, peer-learners' engagement expressed by hand-raising behavior~\cite{hand_raising_classroom_learning} may also affect the attention and visual behaviors of learners in the VR classrooms, which could be further studied.

In summary, while showing that VR could be a useful technology to support education, the aforementioned works primarily focused on the importance of used mediums and configurations, visualization styles, participant locations for visual attention, engagement, motivation, and learning of participants in VR classrooms. Yet, real-time and moment-to-moment interactions with the environment and visual behaviors of students in an everyday VR classroom setup were not studied in depth. Although obtaining such information in real-time is challenging, analyzing eye-gaze and eye movement features can provide valuable understanding into visual attention and interaction in a non-intrusive way, especially for designing such classroom configurations. For instance, long fixations can be related to the increased amount of cognitive process ~\cite{just1976eye}, whereas long saccadic behaviors are related to inefficient search behavior ~\cite{goldberg1999computer}. Furthermore, pupillometry is highly related to cognitive workload ~\cite{appel2018cross,appel2019predicting}. Such information is also argued for consideration in IVR environments ~\cite{bailenson_2002,bailenson_2004}. In fact, when designing immersive VR environments for digital transformations of classrooms in virtual worlds, such features can be key to understand visual attention, cognitive processes, and visual interactions towards different classroom manipulations, which may also affect learning and teaching processes. To address this research gap, we study three configurations in an everyday VR classroom setup including different visualization styles of virtual avatars, sitting positions of participants, and hand-raising based performance levels of peer-learners by using eye movement features.

\begin{figure*}[ht]
\centering
\subfigure[Back sitting participant experiencing the VR classroom.]{
    {\includegraphics[width=0.4\linewidth,keepaspectratio]{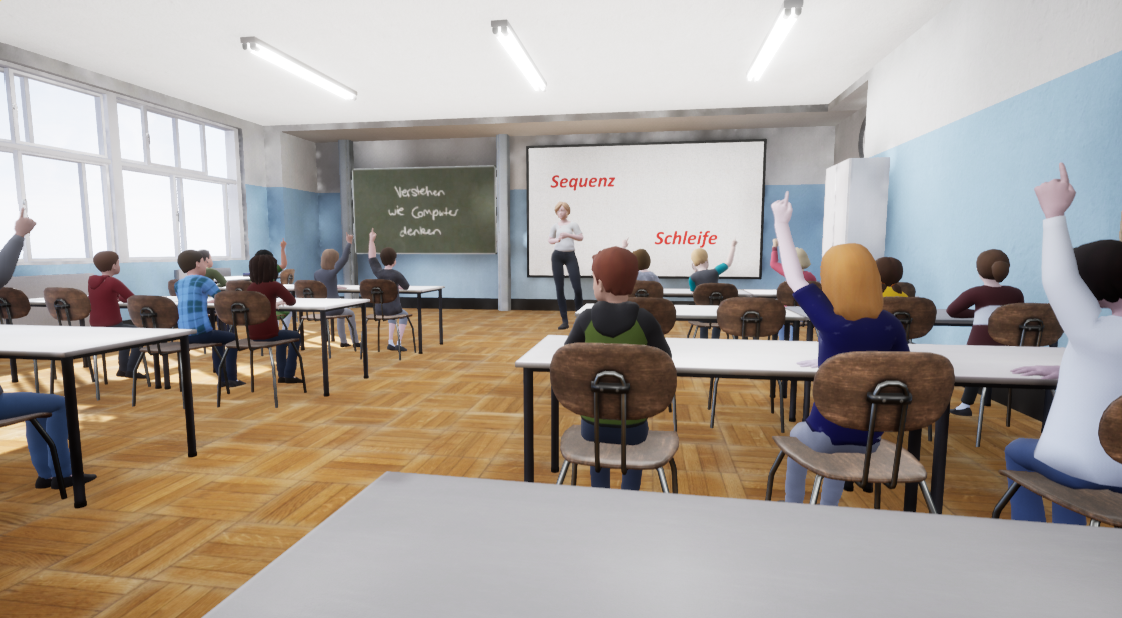}}
}
\qquad
\qquad
\subfigure[Front sitting participant experiencing the VR classroom.]{
    {\includegraphics[width=0.4\linewidth,keepaspectratio]{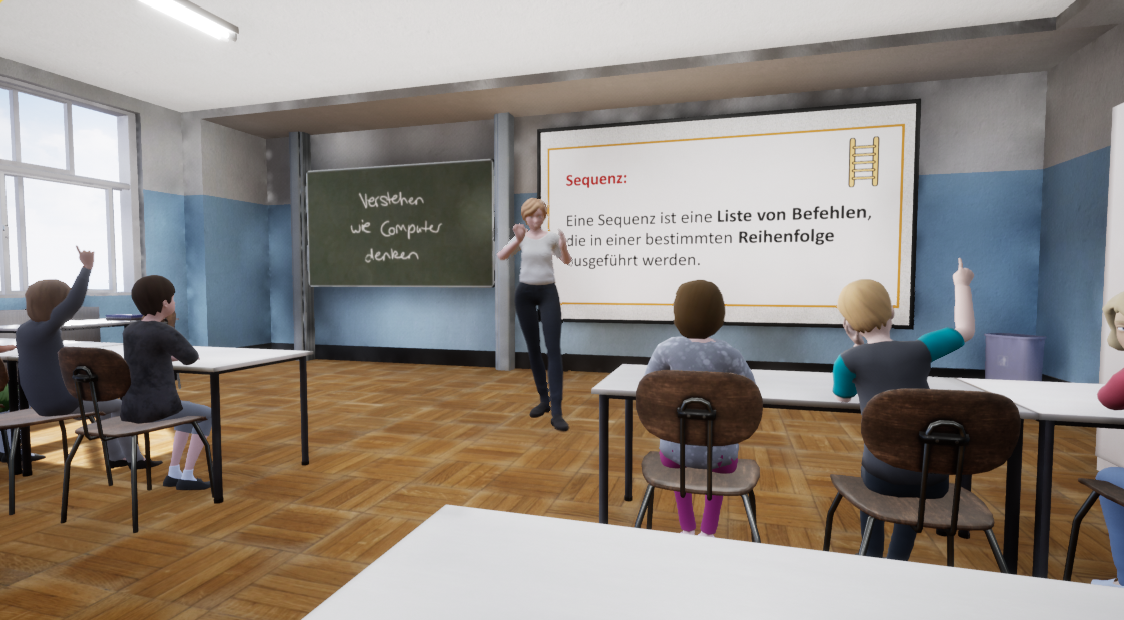}}
}
\qquad
\qquad
\subfigure[Cartoon-styled avatars.]{
    {\includegraphics[width=0.4\linewidth,keepaspectratio]{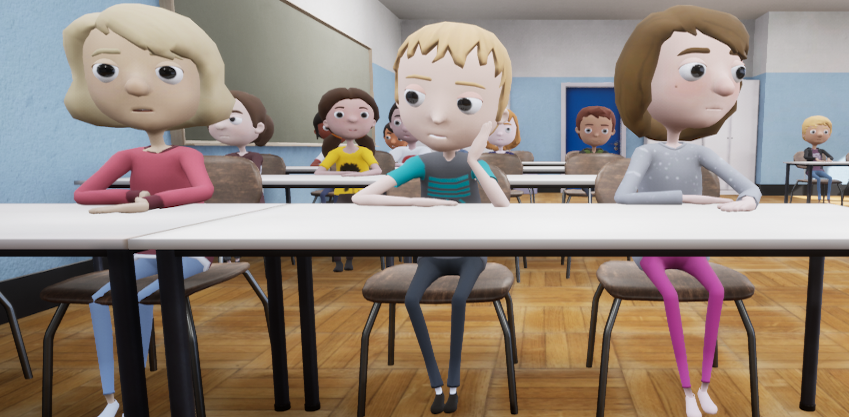}}
}
\qquad
\qquad
\subfigure[Realistic-styled avatars.]{
    {\includegraphics[width=0.4\linewidth,keepaspectratio]{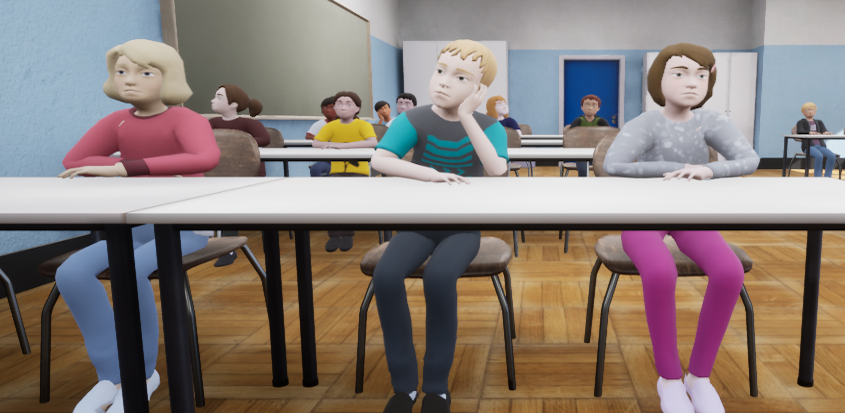} }
}
\caption{Views from the immersive virtual reality classroom.}
\Description{Examples from different parts and configurations of the virtual classroom during learning with four subfigures. The first two subfigures correspond to views from different sitting positions, whereas the latter two show virtual peer-learners with different visualization styles. In the first two subfigures, some of the virtual peer-learners raise their hands for interaction.}
\label{fig:screenshots}
\end{figure*}

\section{Methodology}
The main purpose of our study is to investigate the effects of digital transformations of the classrooms to VR settings on learners. Therefore, we designed a user-study to study these effects. In this section, we discuss the participant information, apparatus, experimental design, experiment procedure, measurements, data pre-processing steps, and our hypotheses. Our study and data collection were approved by the institutional ethics committee at the University of T{\"u}bingen (date of approval: 25/11/2019, file number: A2.5.4-106\_aa) as well as the regional council responsible for educational affairs at the district of T{\"u}bingen.

\subsection{Participants}
\label{subsec:participants}
Participants were recruited from local academic track schools via e-mails and invitation letters. After obtaining written informed consent from both students and their parents or legal guardians, all students who indicated interest were admitted to the study. $381$ volunteer sixth-grade students ($179$ female, $202$ male), whose ages range from $10$ to $13$ ($M=11.51$, $SD=0.56$), were recruited to participate in the experiment. Due to hardware problems or incorrect calibration, data from $32$ participants were removed. In addition, data from $61$ participants were also removed due to eye tracker related issues including low eye tracking ratio (lower than $90\%$). Therefore, data from $288$ participants ($137$ female, $151$ male), whose ages range from $10$ to $13$ ($M = 11.47$, $SD = 0.51$), were used for evaluations. We had $16$ different conditions in the experiment and the average number of participants for each condition was $18$ ($SD = 5.3$). In addition to the actual study and data collection, we successfully piloted both our technical setup and the experimental workflow with $55$ similar aged ($M = 11.35$, $SD = 0.52$) sixth-grade students ($20$ female, $35$ male).

\subsection{Apparatus}
\label{subsec:apparatus}
In our experiments we employed HTC Vive Pro Eye devices with a refresh rate of $90$ Hz and a field of view of $110^{\circ}$. The VR environment was designed and rendered using the Unreal Game Engine\footnote{https://www.unrealengine.com/} v$4$.$23$.$1$. The screen resolution for each eye was set to $1440 \times 1600$. To collect eye movement data, we used the integrated Tobii eye tracker with a $120$ Hz sampling rate and a default calibration with $0.5^{\circ}-1.1^{\circ}$ accuracy.

\begin{table*}[ht]
  \caption{Head and eye movement event identification thresholds.}
  \label{tab:events}
  \begin{tabular}{ccc}
    \toprule
    Event & Conditions for velocity ($v$) & Conditions for duration ($\Delta$)\\
    \midrule
    Stationary HMD & $v_{head} < 7^{\circ}/s$ & -\\
    Fixation & $v_{head} < 7^{\circ}/s$ and $v_{gaze} < 30^{\circ}/s$ & $100ms < \Delta_{fixation} < 500ms$ \\
    Saccade & $v_{gaze} > 60^{\circ}/s$ & $30ms < \Delta_{saccade} < 80ms$\\
  \bottomrule
\end{tabular}
\end{table*}

\subsection{Experimental Design}
\label{subsec:experimental_design}
The virtual classroom designed in our study has $4$ rows and $2$ columns of desks along with chairs, as well as other objects which typically exist in the conventional classrooms such as a board and display. In total, there are $24$ virtual peer-learners sitting on the chairs. A virtual teacher standing in front of the classroom teaches a $\approx 15$-minute virtual lecture to the students about computational thinking~\cite{Weintrop2016DefiningCT}. During the lecture, the virtual teacher walks around the podium. The virtual peer-learners and participants sit on the chairs throughout the lecture. The lecture has four phases including \textbf{(a) topic introduction} ($\approx3$ minutes), \textbf{(b) knowledge input} ($\approx4.5$ minutes), \textbf{(c) exercises} ($\approx5.5$ minutes), and \textbf{(d) summary} ($\approx1.5$ minutes). There are distracting behaviors from virtual peer-learners (e.g., raising hands, turning around) in the first, second, and third phases of the lecture.

In the beginning of the first phase, the teacher enters the classroom, stays in the classroom for a while, and then leaves for $\approx20$ seconds, giving participants the opportunity to look around and adjust to the virtual environment. The topic of the lecture is displayed on the board as \emph{“Understanding how computers think”}. During the first phase, the teacher asks five simple questions to interact with the students. Some of the peer-learners raise their hands and answer the questions. In the second phase, the teacher explains two terms to the students, namely, the terms \emph{``loop''} and \emph{``sequence''}. These terms are also shown on the display. Then, the teacher asks four questions about each term and the peer-learners raise their hands to answer the questions. In the third phase, the teacher gives the students two exercises to evaluate whether or not they understand the terms correctly. For each exercise, the students have some time to think. Then, the teacher provides the answers for each exercise, and the peer-learners vote for the correct answer by raising their hands. In the last phase, the teacher stands in the middle of the classroom to summarize the lecture. No questions are asked in this phase; therefore, none of the peer-learners raise their hands.

Our study is in between-subjects design. The participants are located either in the front or back region of the virtual classroom. The participants that sit in the front of the virtual classroom have one row in front of them, whereas the participants that sit in the back have three rows in front of them. The visualization styles of the avatars have two levels as well, in particular cartoon and realistic. Lastly, the hand-raising percentages, which are intended to show the performance levels of the virtual peer-learners, have four different levels, including $20\%$, $35\%$, $65\%$, and $80\%$. Combining all, we have a $2 \times 2 \times 4$ factorial design that forms $16$ different conditions in total. Participants' views from back and front sitting positions, cartoon- and realistic-styled avatars are depicted in Figures~\ref{fig:screenshots} (a), (b), (c), and (d), respectively.

\subsection{Procedure}
\label{subsec:procedure}
Each experimental session took $\approx45$ minutes including preparation time. We conducted the experiments in groups of ten participants by assigning each participant randomly to one of the sixteen conditions. Before the data assessment took place at the participating schools, students were informed that they could drop out of the study at any time without consequences. After a brief introduction to the experiment and the data collection process, participants had the opportunity to acclimate with the hardware and the VR environment.

The experiment started with the eye tracker calibration. After calibration success, the experimenters pressed the ``Enter'' button to start the actual experiment and data collection process, wherein participants experienced the immersive virtual environment and the lecture. The experiments were supposed to be carried out in one session without breaks, mimicking thus a real classroom teaching session, lasting about 15 minutes. At the end of the experiment, the VR application displayed a message telling the participants to take off their HMDs. Lastly, participants filled out questionnaires about their experienced presence and perceived realism.

\subsection{Measurements}
\label{subsec:measurements}
For this work, our main focus was eye-gaze, head-pose, and pupil related activities of the participants as these are considered to be rich information sources, especially in VR. Fixations are the periods during which eyes are stationary within the head while fixated on an area of interest. Saccades, on the other hand, are the high-speed ballistic eye movements that shift eye-gaze from one fixation to another.

\begin{figure*}
  \centering
   \subfigure[Mean fixation durations.]{{\includegraphics[width=0.315\linewidth,keepaspectratio]{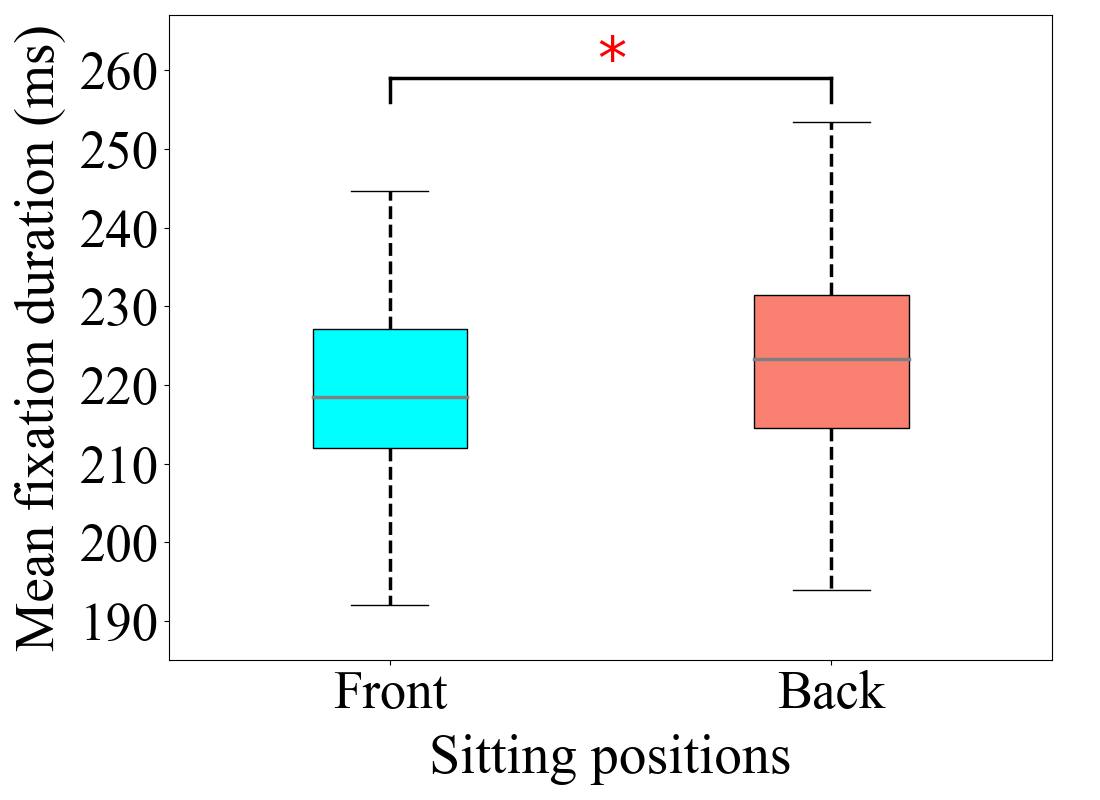}}}%
   \quad
   \subfigure[Mean saccade durations.]{{\includegraphics[width=0.315\linewidth,keepaspectratio]{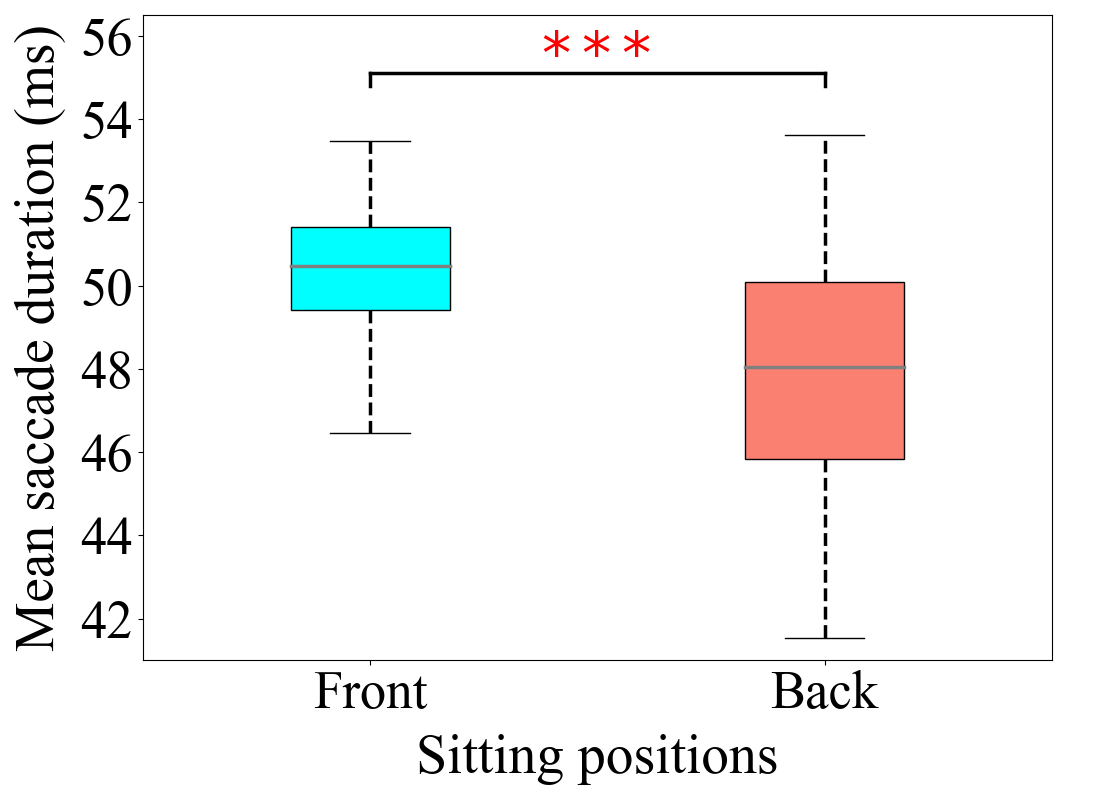} }}%
   \quad
   \subfigure[Mean saccade amplitudes.]{{\includegraphics[width=0.315\linewidth,keepaspectratio]{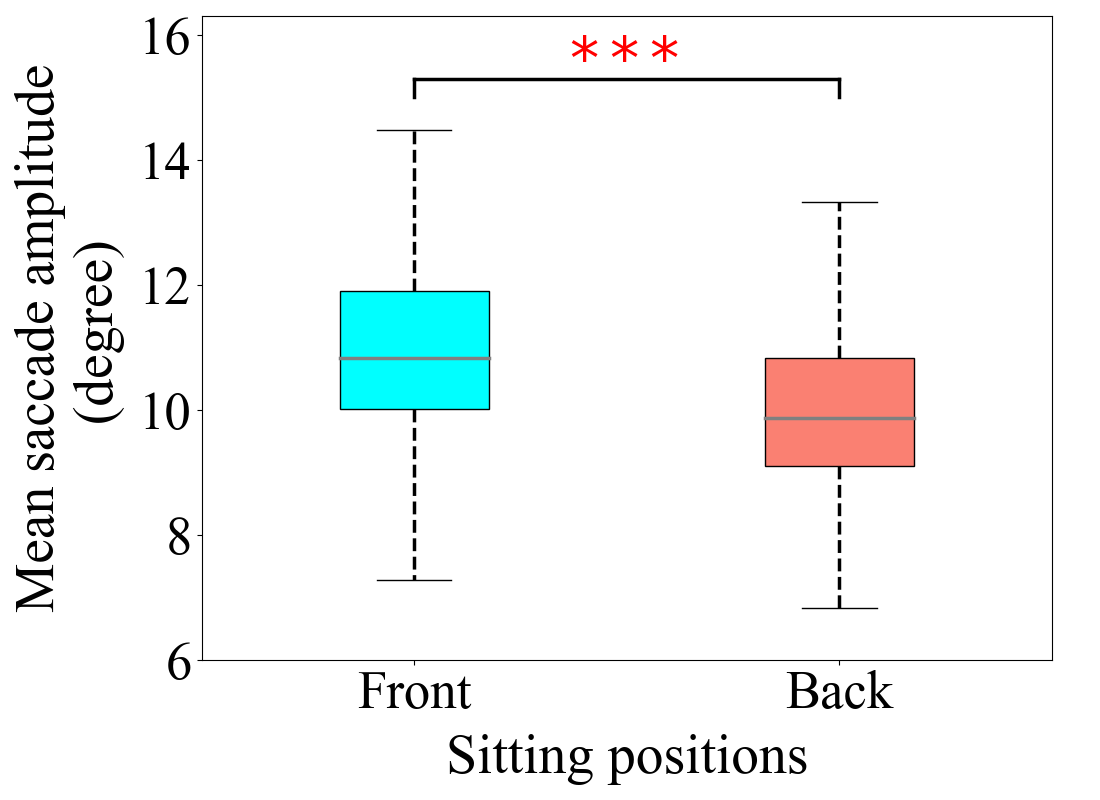}}}
  \caption{Results for different sitting positions. Significant differences are highlighted with * and *** for $p <.05$ and $p <.001$, respectively.}
  \Description{Boxplots of mean fixation durations, mean saccade durations, and mean saccade amplitudes for different sitting position conditions. Horizontal axes correspond to front and back conditions, whereas vertical axes correspond to values of measurements per eye tracking feature in each subfigure.}
  \label{fig:sitting_pos_results}%
\end{figure*}

Using fixations, saccades, and pupil diameters, plenty of eye movement features are extracted. In this study, we extracted the number of fixations, fixation durations, saccade durations, saccade amplitudes, and normalized pupil diameters to analyze different conditions of the experiment. In the eye tracking literature, longer fixation durations correspond to engaging more with the object or increased cognitive process~\cite{just1976eye}. Fixation durations are mainly related to cognition and attention; however, it is argued that they are affected by the procedures that lead to learning and it is reported that fixation durations can be used to understand learning processes as well ~\cite{Negi_Mitra_2020}. For instance, ~\cite{CHIEN2015191} has studied fixation patterns during learning in simulation- and microcomputer-based laboratory and found that simulation group had longer fixation duration, which means more attention and deeper cognitive processing. In addition to the fixations, longer saccade durations correspond to less efficient scanning or searching~\cite{goldberg1999computer}, whereas longer saccade amplitudes mean that attention is drawn from a distance~\cite{goldberg2002eye}. Furthermore, a larger pupil diameter is related to higher cognitive load~\cite{beatty:1982}. In addition, while being task dependent, ~\cite{doi:10.1177/1541931213601689} has indicated that pupil diameter measurements in high task load correlate with individual's performance. However, as pupil diameter values are also affected by the illumination, a controlled environment is needed to assess it. In our VR setup, the illumination is controlled across different conditions. Besides, a general overview of considering eye tracking as a tool to enhance learning with graphics is provided in ~\cite{MAYER2010167}.

Additionally, the self-reported presence and realism were assessed by questionnaires. The items in the questionnaires were based on the conceptualizations of ~\cite{schubert_presence} and ~\cite{lombard_presence} which were developed particularly to assess students' perception of the VR classroom situation. The experienced presence and perceived realism were assessed via using a 4-point Likert scales ranging from 1 (``do not agree at all'') to 4 (``completely agree'') with nine (e.g., ``I felt like I was sitting in the virtual classroom.'' or ``I felt like the teacher in the virtual classroom really addressed me.'') and six items (e.g., ``What I experienced in the virtual classroom, could also happen in a real classroom.'' or ``The students in the virtual classroom behaved similarly to real classmates.''), respectively.

\subsection{Data Pre-processing}
As the raw eye tracking data collected from the VR device does not include fixations, saccades or similar eye movements, we first pre-processed the data to identify these events. Detecting different eye movements in the VR setup is a challenging task and different from the traditional eye tracking experiments that include equipment such as chin-rests, as participants have opportunity to move their heads freely in VR. In the eye tracking literature, Velocity-Threshold Identification (I-VT) method is used to classify fixations based on velocities~\cite{salvucci2000identifying}. In the VR context, ~\cite{agtzidis2019360} applied a similar method to detect eye movement events. We opted for a similar approach.

Before applying the I-VT, we first applied linear interpolation for the missing gaze vectors. After the interpolation, we identified the fixations when the HMD was stationary. However, the identification of saccades was not restricted by the HMD movement. The used velocity and duration thresholds for the HMD movement states, fixations, and saccades are depicted in Table~\ref{tab:events}, where the velocities and durations are given as $v$ and $\Delta$, respectively. Unlike the fixations and saccades, the pupil diameter values are reported by the eye tracker. As raw pupil diameter values are affected by blinks and noisy sensor readings, we smoothed and normalized the pupil diameter readings using Savitzky-Golay filter~\cite{savitzky64} and divisive baseline correction using a baseline duration of $\approx 1$ seconds~\cite{Mathot2018}, respectively.

\subsection{Hypotheses}
We developed three hypotheses, each corresponds to one design factor.

\begin{itemize}
\item \textbf{Hypothesis-1 (H1)}: We hypothesize that the different sitting positions of the participants yield different effects on the eye movements. As the participants that sit in the front are closer to the board, displays, and the teacher, we assume that they can attend the virtual lecture more efficiently than participants in the back and have less difficulty extracting information about the lecture. However, as they have a narrower field of view, particularly towards the frontal part of the classroom, they need to shift their attention more than the participants sitting in the back.

\item \textbf{Hypothesis-2 (H2)}: We hypothesize that different visualization styles of virtual avatars affect student visual behaviors differently. More particularly, as students are familiar with realistic styles in the conventional classrooms, we claim that compared to cartoon-styled visualization condition, they attend the scene shorter during fixations in the realistic-styled visualization setting as cartoon-styled avatars are more attractive to the students. Therefore, students engage with the environment more in the cartoon-styled visualization condition than in the realistic-styled condition.

\item \textbf{Hypothesis-3 (H3)}: We hypothesize that different hand-raising percentages of virtual peer-learners can distinctively affect the behaviors of participants. Specifically, we anticipate that when relatively higher percentages of hand-raising levels are provided, such as $65\%$ or $80\%$, the participant's cognitive load will be higher due to the fact that many of the peer-learners attend the lecture with a high focus. Similarly, participants have more fixations in the classroom in the higher hand-raising percentage conditions as a higher number of hand-raising percentage creates an opportunity for various attention and distraction points.
\end{itemize}

\section{Results}
\label{sec:results}
As we have three factors that form $16$ different conditions, we applied $3$-way full-factorial analysis of variance (ANOVA) by setting the level of significance to $\alpha = 0.05$ with Tukey-Kramer post-hoc test. For the non-parametric factorial analysis, we used the Aligned Rank Transform (ART)~\cite{10.1145/1978942.1978963} before applying ANOVA procedures.

\begin{figure*}
  \centering
   \subfigure[Mean fixation durations.]{{\includegraphics[width=0.315\linewidth,keepaspectratio]{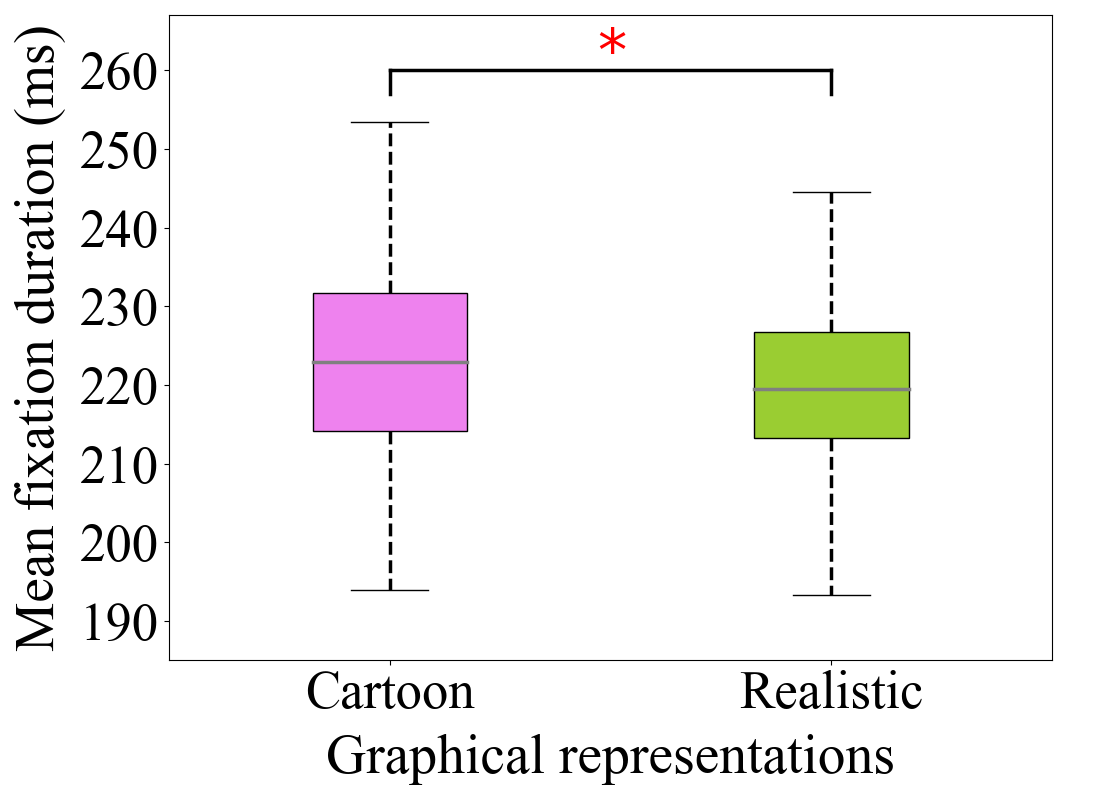}}}%
   \quad
   \subfigure[Mean saccade durations.]{{\includegraphics[width=0.315\linewidth,keepaspectratio]{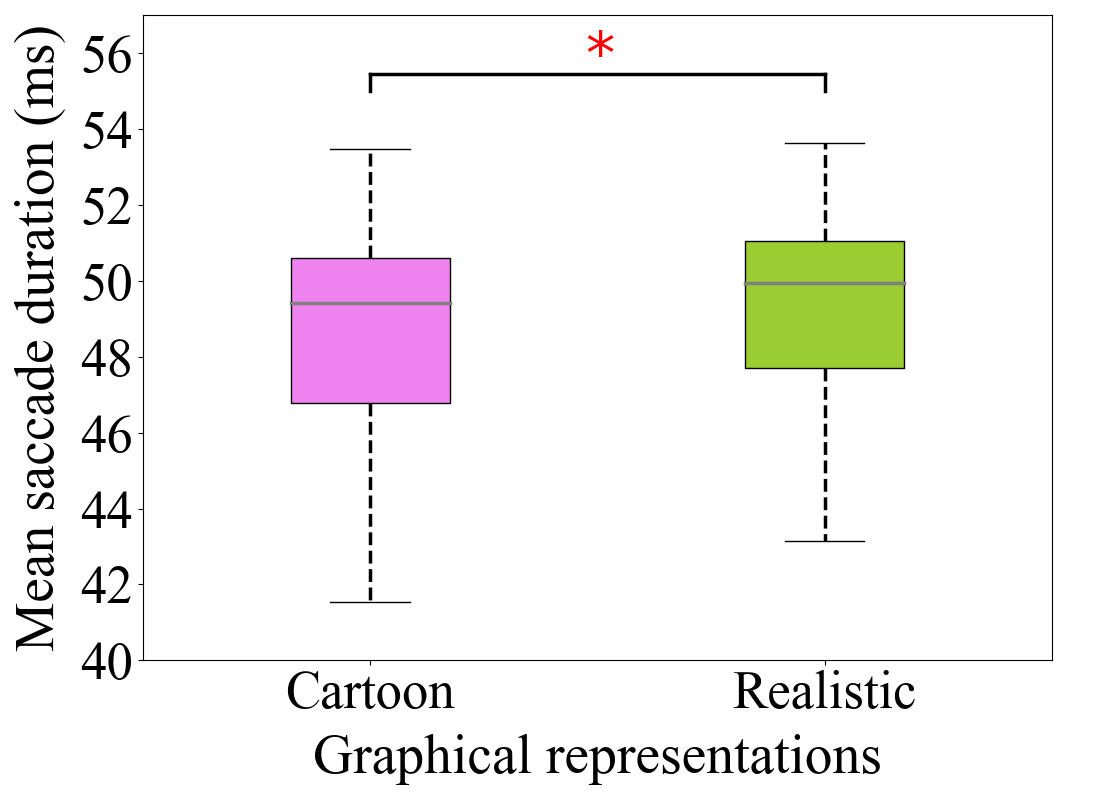} }}%
    \quad
   \subfigure[Pupil diameters.]{{\includegraphics[width=0.315\linewidth,keepaspectratio]{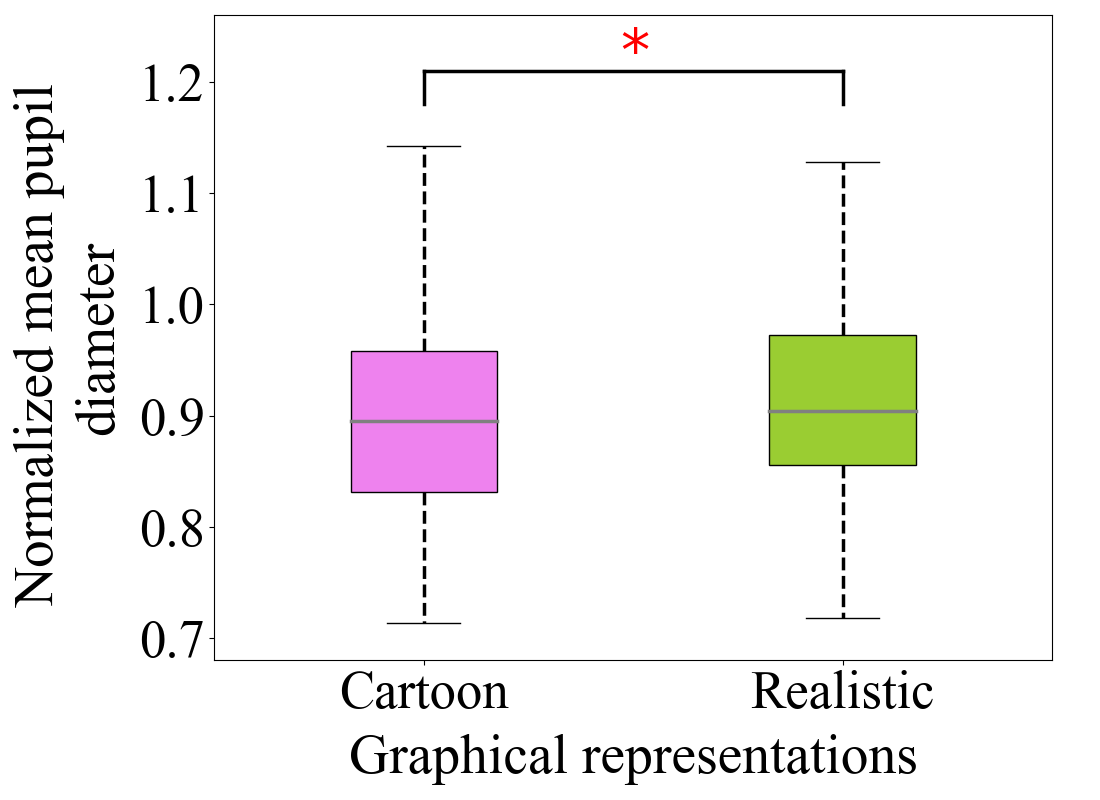}}}
  \caption{Results for different avatar visualization styles. Significant differences are highlighted with * for $p <.05$.}
  \Description{Boxplots of mean fixation durations, mean saccade durations, and normalized mean pupil diameters for different avatar visualization style conditions. Horizontal axes correspond to cartoon and realistic style conditions, whereas vertical axes correspond to values of measurements per eye tracking feature in each subfigure.}
  \label{fig:different_avatar_results}
\end{figure*}

\subsection{Analysis on Different Sitting Positions}
Different sitting positions have an impact on the mean fixation and saccade durations, and mean saccade amplitudes. The mean fixation durations of the front and back sitting participants are illustrated in Figure ~\ref{fig:sitting_pos_results} (a). The participants that sit in the back have significantly longer mean fixation durations ($M = 222.6ms, SD = 14.57ms$) than the participants that sit in the front ($M = 218.75ms, SD = 13.11ms$), with $F(1,272) = 6.7$, $p = .01$.

Both saccade durations and amplitudes are influenced by the sitting positions and are depicted in Figures ~\ref{fig:sitting_pos_results} (b) and (c), respectively. The results reveal significantly longer saccade durations in the front condition ($M = 50.23ms, SD = 1.7ms$) than in the back condition ($M = 47.9ms, SD = 2.62ms$), with $F(1,272) = 73.76$, $p<.001$. Similarly, the mean saccade amplitude is significantly larger in the front condition ($M = 10.93^{\circ}, SD = 1.54^{\circ}$) than in the back condition ($M = 10.05^{\circ}, SD = 1.38^{\circ}$), with $F(1,272) = 22.6$, $p < .001$.

\subsection{Analysis on Different Avatar Styles}
Different avatar visualization styles affect the mean fixation and saccade durations, and pupil diameters. The results are depicted in Figures ~\ref{fig:different_avatar_results} (a), (b), and (c), respectively. The mean fixation durations are significantly longer in the cartoon-styled avatar condition ($M = 222.88ms, SD = 14.06ms$) than in the realistic-styled avatar condition ($M = 218.6ms, SD = 13.76ms$), with $F(1,272) = 5.27$, $p = .022$. By contrast, the mean saccade durations are significantly shorter in the cartoon-styled avatar condition ($M = 48.58ms, SD = 2.66ms$) than in the realistic-styled condition ($M = 49.3ms, SD = 2.35ms$), with $F(1,272) = 6.22$, $p = .013$.

The normalized mean pupil diameter, which reflects the cognitive load, is significantly larger in the realistic-styled avatar condition ($M = 0.94, SD = 0.16$) than in the cartoon-styled avatar condition ($M = 0.91, SD = 0.13$), with $F(1,272) = 3.94$, $p = .048$.

\subsection{Analysis on Different Hand-raising Behaviors}
\begin{figure*}
  \centering
   \subfigure[Pupil diameters.]{{\includegraphics[width=0.485\linewidth,keepaspectratio]{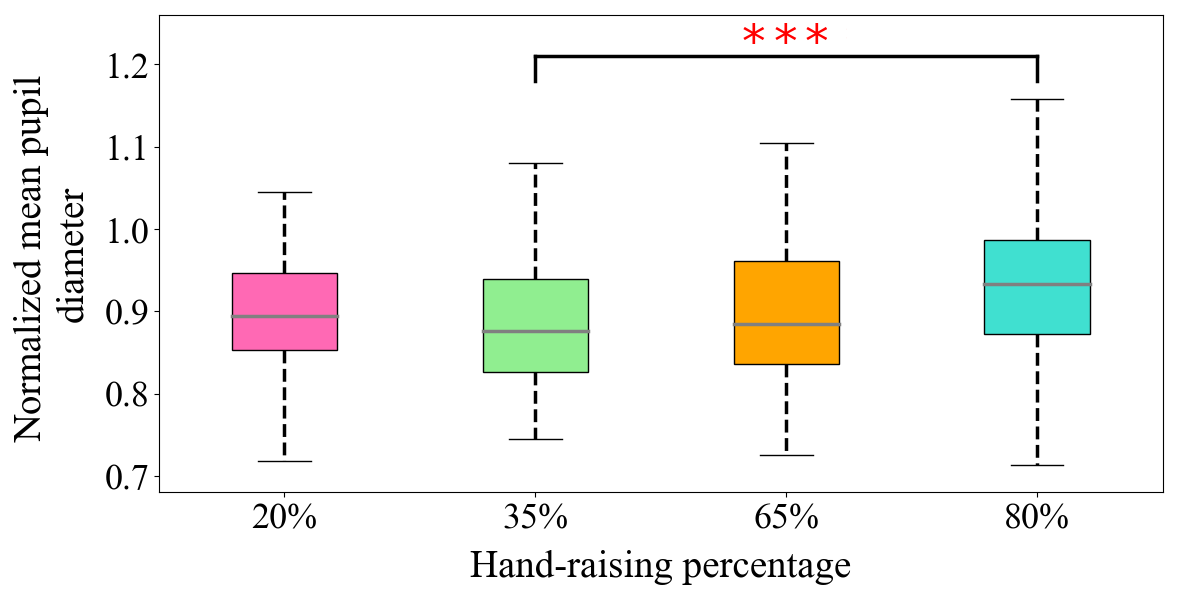} }}%
   \quad
   \subfigure[Number of fixations.]{{\includegraphics[width=0.485\linewidth,keepaspectratio]{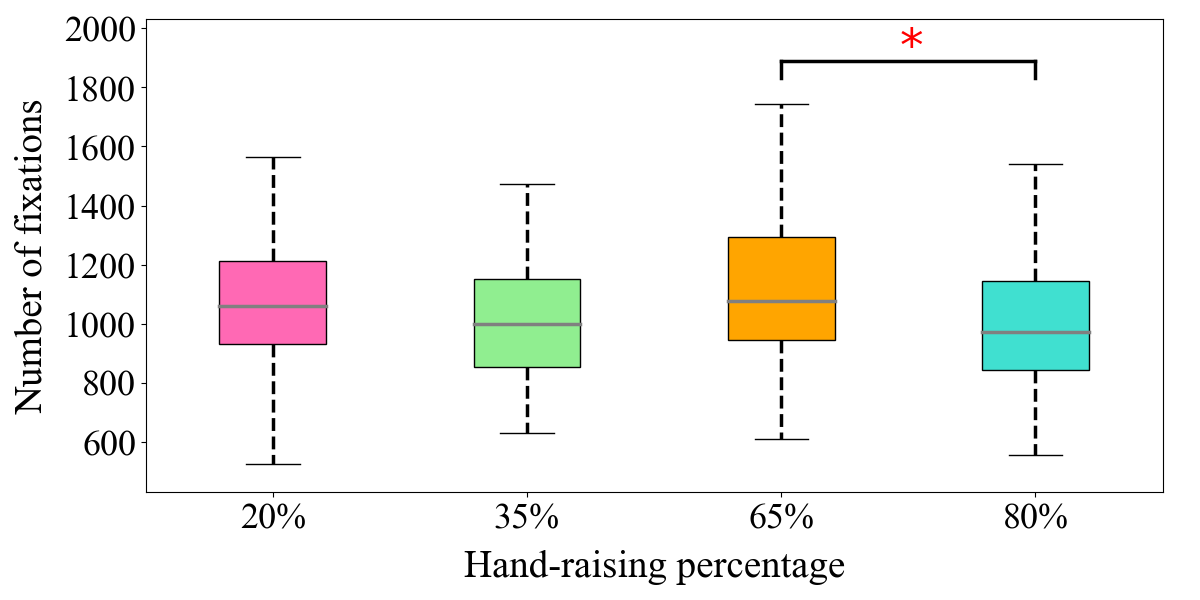}}}%
  \caption{Results for different hand-raising percentages. Significant differences are highlighted with * and *** for $p <.05$ and $p <.001$, respectively.}
  \Description{Boxplots of normalized mean pupil diameters and number of fixations for different hand-raising conditions. Horizontal axes correspond to percentages of hand-raising virtual peer-learners, whereas vertical axes correspond to values of measurements per eye tracking feature in each subfigure.}
  \label{fig:different_handraising_results}%
\end{figure*}

The hand-raising behaviors of virtual peer-learners have significant impacts on the pupil diameters and number of fixations as depicted in Figures ~\ref{fig:different_handraising_results} (a) and (b), respectively. We found significant effects on normalized mean pupil diameter values with $F(3,272) = 4.78$, $p = .003$. Particularly, mean pupil diameter in the $80\%$ hand-raising condition ($M = 0.96, SD = 0.16$) is significantly larger than in the $35\%$ hand-raising condition ($M = 0.9, SD = 0.12$), with $F(3,272) = 4.78$, $p < .001$. In addition, we found significant effects on number of fixations with $F(3,272) = 3.01$, $p = .03$. More specifically, there are notably more fixations in the $65\%$ hand-raising condition ($M = 1112.92, SD = 245.07$) than in the $80\%$ hand-raising condition ($M = 995.49, SD = 211.98$), with $F(3,272) = 3.01$, $p = .028$.

\subsection{Analysis on Experienced Presence and Perceived Realism}
We did not find significant effects of different experimental conditions on the self-reported experienced presence and perceived realism. Overall, the self-reported experienced presence and perceived realism values are in the vicinity of highest values with ($M = 2.91$, $SD = 0.55$) and ($M = 2.91$, $SD = 0.57$), respectively. These mean that even though we did not obtain statistically significant differences between conditions, the participants experienced high levels of presence and realism in the IVR classroom environment.

\section{Discussion}
The results show that there are significant differences in the eye movement features between front and back sitting position conditions. Firstly, participants had longer fixations in the back sitting condition. This indicates that they had more processing time than the participants sitting in the front, which can be related to difficulty extracting information, similar to the relationship between task difficulty and mean fixation duration ~\cite{task_difficulty_fixation_durations}. Secondly, the participants that sit in the front had longer saccade durations and amplitudes, which suggests that they needed to shift their attention more during the virtual lecture. While being located closer to the lecture content, longer saccade durations indicate that the participants sitting in the front had less efficient scanning behavior ~\cite{goldberg1999computer} during the lecture. We assume that this was due to the narrower field of view. These results support our \textbf{H1}. When designing virtual classes, these results should be taken into account, particularly when determining where students should be located in the classroom, depending on the context.

Our results show consequential effects in the eye movement features in different avatar style conditions. As mean fixation durations are longer in the cartoon-styled visualization condition, we assume participants found the cartoon-styled avatars more attractive and attention-grabbing. Therefore, their fixation behaviors were longer during the virtual lecture. On the contrary, the mean saccade durations are longer in realistic-styled conditions as the fixation durations are shorter, which is theoretically expected. Furthermore, the pupil diameters of the participants in the realistic-styled condition are larger, indicating that the cognitive load of these participants was significantly higher during the lecture, which is suggested by the previous work ~\cite{beatty:1982}. This is an indication that participants may have taken the lecture more seriously and in a more focused manner when the visualization was realistic. These findings support our \textbf{H2}. Rendering realistic-styled avatars may be computationally expensive depending on the configuration. Therefore, an optimal trade-off should be decided, taking the behavioral results into account while designing the virtual classrooms.

Furthermore, we observe significant effects in attention towards different hand-raising based performance levels of the peer-learners. Particularly, the pupil diameters of the participants in the $80\%$ condition are significantly larger than the pupil diameters of the participants in the $35\%$ condition. We interpret this to mean that when the performance and attendance level of peer-learners was relatively higher, the participants' cognitive load became higher, indicating that they might pay more attention to the lecture content. This partially supports our \textbf{H3}. In addition, a greater number of fixations are observed in the $65\%$ condition than in the $80\%$ condition. We claim that when almost all of the peer-learners participated in hand-raising behaviors during the lecture, participants acknowledged this information without significantly shifting their gaze. However, this claim requires further investigation. Manipulation of different hand-raising conditions may affect student self-concept~\cite{self_concept_1976}, which should be further studied as well.

In our study, the interaction and perception in the immersive VR classroom were assessed mainly by using eye-gaze and head-pose information. However, while the virtual teacher and peer-learners talk in the simulations, no response or interaction by means of audio or gestures was expected from the participants. Combining visual perceptions and interactions with such data may provide additional insights particularly for better interaction design in VR classrooms. A future iteration can also evolve into an everyday virtual classroom platform where each virtual agent is actually connected to a real person, similar to in platforms such as Mozilla Hubs. To this end, further design settings such as optimal seating arrangement (e.g., U-shape, circle shape) in addition to the sitting positions should be investigated. Evaluation of similar configurations in online learning platforms such as Coursera\footnote{https://www.coursera.org/}, Udemy\footnote{https://www.udemy.com/}, or MOOCs\footnote{https://www.mooc.org/} could provide additional implications for interaction modeling. Furthermore, gaze-based attention guidance can be considered for more interactive VR classroom experience and it can be achieved by fine-grained eye movement analysis focusing on short time windows instead of complete experiments. While being out of the scope of this paper, assessing learning outcomes and combining them with visual interaction and scanpath behaviors from immersive VR classroom could also offer insights for optimal VR classroom design.

\section{Conclusion}
In this work, we evaluated three major design factors of immersive VR classrooms, namely different participant locations in the virtual classroom, different visualization styles of virtual peer-learners and teachers, including cartoon and realistic, and different hand-raising behaviors of peer-learners, particularly through the analysis of eye tracking data. Our results indicate that participants located in the back of the virtual classroom may have difficulty extracting information during the lecture. In addition, if the avatars in the classroom are visualized in realistic styles, participants may attend the lecture in a more focused manner instead of being distracted by the visualization styles of the avatars. These findings offer valuable insights about design decisions in the VR classroom environment. Few indicators were obtained from the evaluation of the different hand-raising behaviors of peer-learners, providing a general understanding of attention towards peer-learner performance. However, these indicators should be further investigated and remain a focus of future work. 

\begin{acks}
This research was partly supported by a grant to Richard G{\"o}llner funded by the Ministry of Science, Research and the Arts of the state of Baden-W{\"u}rttemberg and the University of T{\"u}bingen as part of the Promotion Program of Junior Researchers. Lisa Hasenbein is a doctoral candidate and supported by the LEAD Graduate School \& Research Network, which is funded by the Ministry of Science, Research and the Arts of the state of Baden-W{\"u}rttemberg within the framework of the sustainability funding for the projects of the Excellence Initiative II. Authors thank Stephan Soller, Sandra Hahn, and Sophie Fink from the Hochschule der Medien Stuttgart for their work and support related to the immersive virtual reality classroom used in this study.
\end{acks}

\bibliographystyle{ACM-Reference-Format}
\bibliography{references}


\begin{thebibliography}{62}


\ifx \showCODEN    \undefined \def \showCODEN     #1{\unskip}     \fi
\ifx \showDOI      \undefined \def \showDOI       #1{#1}\fi
\ifx \showISBNx    \undefined \def \showISBNx     #1{\unskip}     \fi
\ifx \showISBNxiii \undefined \def \showISBNxiii  #1{\unskip}     \fi
\ifx \showISSN     \undefined \def \showISSN      #1{\unskip}     \fi
\ifx \showLCCN     \undefined \def \showLCCN      #1{\unskip}     \fi
\ifx \shownote     \undefined \def \shownote      #1{#1}          \fi
\ifx \showarticletitle \undefined \def \showarticletitle #1{#1}   \fi
\ifx \showURL      \undefined \def \showURL       {\relax}        \fi
\providecommand\bibfield[2]{#2}
\providecommand\bibinfo[2]{#2}
\providecommand\natexlab[1]{#1}
\providecommand\showeprint[2][]{arXiv:#2}

\bibitem[\protect\citeauthoryear{Agtzidis, Startsev, and Dorr}{Agtzidis
  et~al\mbox{.}}{2019}]%
        {agtzidis2019360}
\bibfield{author}{\bibinfo{person}{Ioannis Agtzidis}, \bibinfo{person}{Mikhail
  Startsev}, {and} \bibinfo{person}{Michael Dorr}.}
  \bibinfo{year}{2019}\natexlab{}.
\newblock \showarticletitle{360-Degree Video Gaze Behaviour: A Ground-Truth
  Data Set and a Classification Algorithm for Eye Movements}. In
  \bibinfo{booktitle}{\emph{Proceedings of the 27th ACM International
  Conference on Multimedia}} (Nice, France). \bibinfo{publisher}{ACM},
  \bibinfo{address}{New York, NY, USA}, \bibinfo{pages}{1007–1015}.
\newblock
\showISBNx{9781450368896}
\urldef\tempurl%
\url{https://doi.org/10.1145/3343031.3350947}
\showDOI{\tempurl}


\bibitem[\protect\citeauthoryear{Alhalabi}{Alhalabi}{2016}]%
        {alhalabi2016virtual}
\bibfield{author}{\bibinfo{person}{Wadee Alhalabi}.}
  \bibinfo{year}{2016}\natexlab{}.
\newblock \showarticletitle{Virtual reality systems enhance students’
  achievements in engineering education}.
\newblock \bibinfo{journal}{\emph{Behaviour \& Information Technology}}
  \bibinfo{volume}{35} (\bibinfo{date}{07} \bibinfo{year}{2016}),
  \bibinfo{pages}{1--7}.
\newblock
\urldef\tempurl%
\url{https://doi.org/10.1080/0144929X.2016.1212931}
\showDOI{\tempurl}


\bibitem[\protect\citeauthoryear{Appel, Scharinger, Gerjets, and Kasneci}{Appel
  et~al\mbox{.}}{2018}]%
        {appel2018cross}
\bibfield{author}{\bibinfo{person}{Tobias Appel}, \bibinfo{person}{Christian
  Scharinger}, \bibinfo{person}{Peter Gerjets}, {and}
  \bibinfo{person}{Enkelejda Kasneci}.} \bibinfo{year}{2018}\natexlab{}.
\newblock \showarticletitle{Cross-subject workload classification using
  pupil-related measures}. In \bibinfo{booktitle}{\emph{Proceedings of the 2018
  ACM Symposium on Eye Tracking Research \& Applications}} (Warsaw, Poland).
  \bibinfo{publisher}{ACM}, \bibinfo{address}{New York, NY, USA}, Article
  \bibinfo{articleno}{4}, \bibinfo{numpages}{8}~pages.
\newblock
\urldef\tempurl%
\url{https://doi.org/10.1145/3204493.3204531}
\showDOI{\tempurl}


\bibitem[\protect\citeauthoryear{Appel, Sevcenko, Wortha, Tsarava, Moeller,
  Ninaus, Kasneci, and Gerjets}{Appel et~al\mbox{.}}{2019}]%
        {appel2019predicting}
\bibfield{author}{\bibinfo{person}{Tobias Appel}, \bibinfo{person}{Natalia
  Sevcenko}, \bibinfo{person}{Franz Wortha}, \bibinfo{person}{Katerina
  Tsarava}, \bibinfo{person}{Korbinian Moeller}, \bibinfo{person}{Manuel
  Ninaus}, \bibinfo{person}{Enkelejda Kasneci}, {and} \bibinfo{person}{Peter
  Gerjets}.} \bibinfo{year}{2019}\natexlab{}.
\newblock \showarticletitle{Predicting Cognitive Load in an Emergency
  Simulation Based on Behavioral and Physiological Measures}. In
  \bibinfo{booktitle}{\emph{2019 International Conference on Multimodal
  Interaction}} (Suzhou, China). \bibinfo{publisher}{ACM},
  \bibinfo{address}{New York, NY, USA}, \bibinfo{pages}{154--163}.
\newblock
\urldef\tempurl%
\url{https://doi.org/10.1145/3340555.3353735}
\showDOI{\tempurl}


\bibitem[\protect\citeauthoryear{Bailenson, Aharoni, Beall, Guadagno, Dimov,
  and Blascovich}{Bailenson et~al\mbox{.}}{2004}]%
        {bailenson_2004}
\bibfield{author}{\bibinfo{person}{Jeremy~N. Bailenson}, \bibinfo{person}{Eyal
  Aharoni}, \bibinfo{person}{Andrew~C. Beall}, \bibinfo{person}{Rosanna~E.
  Guadagno}, \bibinfo{person}{Aleksandar Dimov}, {and} \bibinfo{person}{Jim
  Blascovich}.} \bibinfo{year}{2004}\natexlab{}.
\newblock \showarticletitle{Comparing behavioral and self-report measures of
  embodied agents' social presence in immersive virtual environments}. In
  \bibinfo{booktitle}{\emph{Proceedings of the 7th Annual International
  Workshop on Presence}}. \bibinfo{publisher}{The International Society for
  Presence Research}, \bibinfo{address}{Valencia, Spain},
  \bibinfo{pages}{216--223}.
\newblock


\bibitem[\protect\citeauthoryear{Bailenson, Beall, and Blascovich}{Bailenson
  et~al\mbox{.}}{2002}]%
        {bailenson_2002}
\bibfield{author}{\bibinfo{person}{Jeremy~N. Bailenson},
  \bibinfo{person}{Andrew~C. Beall}, {and} \bibinfo{person}{Jim Blascovich}.}
  \bibinfo{year}{2002}\natexlab{}.
\newblock \showarticletitle{Gaze and task performance in shared virtual
  environments}.
\newblock \bibinfo{journal}{\emph{The Journal of Visualization and Computer
  Animation}} \bibinfo{volume}{13}, \bibinfo{number}{5} (\bibinfo{year}{2002}),
  \bibinfo{pages}{313--320}.
\newblock
\urldef\tempurl%
\url{https://doi.org/10.1002/vis.297}
\showDOI{\tempurl}


\bibitem[\protect\citeauthoryear{Bailenson, Yee, Blascovich, Beall, Lundblad,
  and Jin}{Bailenson et~al\mbox{.}}{2008}]%
        {bailenson_et_al_2008}
\bibfield{author}{\bibinfo{person}{Jeremy~N. Bailenson}, \bibinfo{person}{Nick
  Yee}, \bibinfo{person}{Jim Blascovich}, \bibinfo{person}{Andrew~C. Beall},
  \bibinfo{person}{Nicole Lundblad}, {and} \bibinfo{person}{Michael Jin}.}
  \bibinfo{year}{2008}\natexlab{}.
\newblock \showarticletitle{The Use of Immersive Virtual Reality in the
  Learning Sciences: Digital Transformations of Teachers, Students, and Social
  Context}.
\newblock \bibinfo{journal}{\emph{Journal of the Learning Sciences}}
  \bibinfo{volume}{17} (\bibinfo{year}{2008}), \bibinfo{pages}{102--141}.
\newblock
\urldef\tempurl%
\url{https://doi.org/10.1080/10508400701793141}
\showDOI{\tempurl}


\bibitem[\protect\citeauthoryear{Beatty}{Beatty}{1982}]%
        {beatty:1982}
\bibfield{author}{\bibinfo{person}{Jackson Beatty}.}
  \bibinfo{year}{1982}\natexlab{}.
\newblock \showarticletitle{Task-evoked pupillary responses, processing load,
  and the structure of processing resources}.
\newblock \bibinfo{journal}{\emph{Psychological Bulletin}}
  \bibinfo{volume}{91}, \bibinfo{number}{2} (\bibinfo{year}{1982}),
  \bibinfo{pages}{276--292}.
\newblock
\urldef\tempurl%
\url{https://doi.org/10.1037/0033-2909.91.2.276}
\showDOI{\tempurl}


\bibitem[\protect\citeauthoryear{Blume, G{\"o}llner, Moeller, Dresler, Ehlis,
  and Gawrilow}{Blume et~al\mbox{.}}{2019}]%
        {blume_et_al_18}
\bibfield{author}{\bibinfo{person}{Friederike Blume}, \bibinfo{person}{Richard
  G{\"o}llner}, \bibinfo{person}{Korbinian Moeller}, \bibinfo{person}{Thomas
  Dresler}, \bibinfo{person}{Ann-Christine Ehlis}, {and}
  \bibinfo{person}{Caterina Gawrilow}.} \bibinfo{year}{2019}\natexlab{}.
\newblock \showarticletitle{Do students learn better when seated close to the
  teacher? A virtual classroom study considering individual levels of
  inattention and hyperactivity-impulsivity}.
\newblock \bibinfo{journal}{\emph{Learning and Instruction}}
  \bibinfo{volume}{61} (\bibinfo{year}{2019}), \bibinfo{pages}{138--147}.
\newblock
\showISSN{0959-4752}
\urldef\tempurl%
\url{https://doi.org/10.1016/j.learninstruc.2018.10.004}
\showDOI{\tempurl}


\bibitem[\protect\citeauthoryear{B{\"o}heim, Urdan, Knogler, and
  Seidel}{B{\"o}heim et~al\mbox{.}}{2020}]%
        {hand_raising_classroom_learning}
\bibfield{author}{\bibinfo{person}{Ricardo B{\"o}heim}, \bibinfo{person}{Tim
  Urdan}, \bibinfo{person}{Maximilian Knogler}, {and} \bibinfo{person}{Tina
  Seidel}.} \bibinfo{year}{2020}\natexlab{}.
\newblock \showarticletitle{Student hand-raising as an indicator of behavioral
  engagement and its role in classroom learning}.
\newblock \bibinfo{journal}{\emph{Contemporary Educational Psychology}}
  \bibinfo{volume}{62} (\bibinfo{year}{2020}), \bibinfo{pages}{101894}.
\newblock
\showISSN{0361-476X}
\urldef\tempurl%
\url{https://doi.org/10.1016/j.cedpsych.2020.101894}
\showDOI{\tempurl}


\bibitem[\protect\citeauthoryear{Bozkir, Geisler, and Kasneci}{Bozkir
  et~al\mbox{.}}{2019a}]%
        {Bozkir}
\bibfield{author}{\bibinfo{person}{Efe Bozkir}, \bibinfo{person}{David
  Geisler}, {and} \bibinfo{person}{Enkelejda Kasneci}.}
  \bibinfo{year}{2019}\natexlab{a}.
\newblock \showarticletitle{Assessment of Driver Attention during a Safety
  Critical Situation in {VR} to Generate {VR}-Based Training}. In
  \bibinfo{booktitle}{\emph{ACM Symposium on Applied Perception 2019}}
  (Barcelona, Spain). \bibinfo{publisher}{ACM}, \bibinfo{address}{New York, NY,
  USA}, Article \bibinfo{articleno}{23}, \bibinfo{numpages}{5}~pages.
\newblock
\showISBNx{9781450368902}
\urldef\tempurl%
\url{https://doi.org/10.1145/3343036.3343138}
\showDOI{\tempurl}


\bibitem[\protect\citeauthoryear{Bozkir, Geisler, and Kasneci}{Bozkir
  et~al\mbox{.}}{2019b}]%
        {bozkir2019person}
\bibfield{author}{\bibinfo{person}{Efe Bozkir}, \bibinfo{person}{David
  Geisler}, {and} \bibinfo{person}{Enkelejda Kasneci}.}
  \bibinfo{year}{2019}\natexlab{b}.
\newblock \showarticletitle{Person Independent, Privacy Preserving, and Real
  Time Assessment of Cognitive Load using Eye Tracking in a Virtual Reality
  Setup}. In \bibinfo{booktitle}{\emph{2019 IEEE Conference on Virtual Reality
  and 3D User Interfaces (VR)}} (Osaka, Japan). \bibinfo{publisher}{IEEE},
  \bibinfo{address}{New York, NY, USA}, \bibinfo{pages}{1834--1837}.
\newblock
\urldef\tempurl%
\url{https://doi.org/10.1109/VR.2019.8797758}
\showDOI{\tempurl}


\bibitem[\protect\citeauthoryear{Casu, Spano, Sorrentino, and Scateni}{Casu
  et~al\mbox{.}}{2015}]%
        {riftart}
\bibfield{author}{\bibinfo{person}{Andrea Casu}, \bibinfo{person}{Lucio~Davide
  Spano}, \bibinfo{person}{Fabio Sorrentino}, {and} \bibinfo{person}{Riccardo
  Scateni}.} \bibinfo{year}{2015}\natexlab{}.
\newblock \showarticletitle{RiftArt: Bringing Masterpieces in the Classroom
  through Immersive Virtual Reality}. In \bibinfo{booktitle}{\emph{Smart Tools
  and Apps for Graphics - Eurographics Italian Chapter Conference}} (Verona,
  Italy). \bibinfo{publisher}{The Eurographics Association},
  \bibinfo{address}{Geneva, Switzerland}, \bibinfo{pages}{77--84}.
\newblock
\showISBNx{978-3-905674-97-2}
\urldef\tempurl%
\url{https://doi.org/10.2312/stag.20151294}
\showDOI{\tempurl}


\bibitem[\protect\citeauthoryear{Cheng and Tsai}{Cheng and Tsai}{2019}]%
        {207ed}
\bibfield{author}{\bibinfo{person}{{Kun Hung} Cheng} {and}
  \bibinfo{person}{{Chin Chung} Tsai}.} \bibinfo{year}{2019}\natexlab{}.
\newblock \showarticletitle{A case study of immersive virtual field trips in an
  elementary classroom: Students{\textquoteright} learning experience and
  teacher-student interaction behaviors}.
\newblock \bibinfo{journal}{\emph{Computers \& Education}}
  \bibinfo{volume}{140} (\bibinfo{year}{2019}), \bibinfo{pages}{103600}.
\newblock
\showISSN{0360-1315}
\urldef\tempurl%
\url{https://doi.org/10.1016/j.compedu.2019.103600}
\showDOI{\tempurl}


\bibitem[\protect\citeauthoryear{Chien, Tsai, Chen, Chang, and Chen}{Chien
  et~al\mbox{.}}{2015}]%
        {CHIEN2015191}
\bibfield{author}{\bibinfo{person}{Kuei-Pin Chien}, \bibinfo{person}{Cheng-Yue
  Tsai}, \bibinfo{person}{Hsiu-Ling Chen}, \bibinfo{person}{Wen-Hua Chang},
  {and} \bibinfo{person}{Sufen Chen}.} \bibinfo{year}{2015}\natexlab{}.
\newblock \showarticletitle{Learning differences and eye fixation patterns in
  virtual and physical science laboratories}.
\newblock \bibinfo{journal}{\emph{Computers \& Education}}
  \bibinfo{volume}{82} (\bibinfo{year}{2015}), \bibinfo{pages}{191--201}.
\newblock
\showISSN{0360-1315}
\urldef\tempurl%
\url{https://doi.org/10.1016/j.compedu.2014.11.023}
\showDOI{\tempurl}


\bibitem[\protect\citeauthoryear{Coyne, Foroughi, and Sibley}{Coyne
  et~al\mbox{.}}{2017}]%
        {doi:10.1177/1541931213601689}
\bibfield{author}{\bibinfo{person}{Joseph~T. Coyne}, \bibinfo{person}{Cyrus
  Foroughi}, {and} \bibinfo{person}{Ciara Sibley}.}
  \bibinfo{year}{2017}\natexlab{}.
\newblock \showarticletitle{Pupil Diameter and Performance in a Supervisory
  Control Task: A Measure of Effort or Individual Differences?}
\newblock \bibinfo{journal}{\emph{Proceedings of the Human Factors and
  Ergonomics Society Annual Meeting}} \bibinfo{volume}{61}, \bibinfo{number}{1}
  (\bibinfo{year}{2017}), \bibinfo{pages}{865--869}.
\newblock
\urldef\tempurl%
\url{https://doi.org/10.1177/1541931213601689}
\showDOI{\tempurl}


\bibitem[\protect\citeauthoryear{D{\'i}az-Orueta, Garc{\'i}a-L{\'o}pez,
  Crespo-Egu{\'i}laz, S{\'a}nchez-Carpintero, Climent, and
  Narbona}{D{\'i}az-Orueta et~al\mbox{.}}{2014}]%
        {DazOrueta2014AULAVR}
\bibfield{author}{\bibinfo{person}{Unai D{\'i}az-Orueta},
  \bibinfo{person}{Cristina Garc{\'i}a-L{\'o}pez}, \bibinfo{person}{Nerea
  Crespo-Egu{\'i}laz}, \bibinfo{person}{Roc{\'i}o S{\'a}nchez-Carpintero},
  \bibinfo{person}{Gema Climent}, {and} \bibinfo{person}{Juan Narbona}.}
  \bibinfo{year}{2014}\natexlab{}.
\newblock \showarticletitle{{AULA} virtual reality test as an attention
  measure: Convergent validity with Conners' Continuous Performance Test}.
\newblock \bibinfo{journal}{\emph{Child Neuropsychology}}  \bibinfo{volume}{20}
  (\bibinfo{year}{2014}), \bibinfo{pages}{328--342}.
\newblock
\urldef\tempurl%
\url{https://doi.org/10.1080/09297049.2013.792332}
\showDOI{\tempurl}


\bibitem[\protect\citeauthoryear{Freina and Ott}{Freina and Ott}{2015}]%
        {review_ivr_education}
\bibfield{author}{\bibinfo{person}{Laura Freina} {and} \bibinfo{person}{Michela
  Ott}.} \bibinfo{year}{2015}\natexlab{}.
\newblock \showarticletitle{A literature review on immersive virtual reality in
  education: State of the art and perspectives}. In
  \bibinfo{booktitle}{\emph{Proceedings of the 11th International Scientific
  Conference eLearning and Software for Education}} (Bucharest, Romania).
  \bibinfo{publisher}{Carol I NDU Publishing House},
  \bibinfo{address}{Romania}, \bibinfo{pages}{133--141}.
\newblock
\urldef\tempurl%
\url{https://doi.org/10.12753/2066-026X-15-020}
\showDOI{\tempurl}


\bibitem[\protect\citeauthoryear{Goldberg and Kotval}{Goldberg and
  Kotval}{1999}]%
        {goldberg1999computer}
\bibfield{author}{\bibinfo{person}{Joseph~H. Goldberg} {and}
  \bibinfo{person}{Xerxes~P. Kotval}.} \bibinfo{year}{1999}\natexlab{}.
\newblock \showarticletitle{Computer interface evaluation using eye movements:
  methods and constructs}.
\newblock \bibinfo{journal}{\emph{International Journal of Industrial
  Ergonomics}} \bibinfo{volume}{24}, \bibinfo{number}{6}
  (\bibinfo{year}{1999}), \bibinfo{pages}{631--645}.
\newblock
\showISSN{0169-8141}
\urldef\tempurl%
\url{https://doi.org/10.1016/S0169-8141(98)00068-7}
\showDOI{\tempurl}


\bibitem[\protect\citeauthoryear{Goldberg, Stimson, Lewenstein, Scott, and
  Wichansky}{Goldberg et~al\mbox{.}}{2002}]%
        {goldberg2002eye}
\bibfield{author}{\bibinfo{person}{Joseph~H. Goldberg},
  \bibinfo{person}{Mark~J. Stimson}, \bibinfo{person}{Marion Lewenstein},
  \bibinfo{person}{Neil Scott}, {and} \bibinfo{person}{Anna~M. Wichansky}.}
  \bibinfo{year}{2002}\natexlab{}.
\newblock \showarticletitle{Eye Tracking in Web Search Tasks: Design
  Implications}. In \bibinfo{booktitle}{\emph{Proceedings of the 2002 Symposium
  on Eye Tracking Research \& Applications}} (New Orleans, LA, USA).
  \bibinfo{publisher}{ACM}, \bibinfo{address}{New York, NY, USA},
  \bibinfo{pages}{51–58}.
\newblock
\showISBNx{1581134673}
\urldef\tempurl%
\url{https://doi.org/10.1145/507072.507082}
\showDOI{\tempurl}


\bibitem[\protect\citeauthoryear{Goldberg, S{\"u}mer, St{\"u}rmer, Wagner,
  G{\"o}llner, Gerjets, Kasneci, and Trautwein}{Goldberg et~al\mbox{.}}{2019}]%
        {goldberg2019attentive}
\bibfield{author}{\bibinfo{person}{Patricia Goldberg},
  \bibinfo{person}{{\"O}mer S{\"u}mer}, \bibinfo{person}{Kathleen St{\"u}rmer},
  \bibinfo{person}{Wolfgang Wagner}, \bibinfo{person}{Richard G{\"o}llner},
  \bibinfo{person}{Peter Gerjets}, \bibinfo{person}{Enkelejda Kasneci}, {and}
  \bibinfo{person}{Ulrich Trautwein}.} \bibinfo{year}{2019}\natexlab{}.
\newblock \showarticletitle{Attentive or Not? Toward a Machine Learning
  Approach to Assessing Students’ Visible Engagement in Classroom
  Instruction}.
\newblock \bibinfo{journal}{\emph{Educational Psychology Review}}
  \bibinfo{volume}{31}, \bibinfo{number}{4} (\bibinfo{year}{2019}),
  \bibinfo{pages}{1--23}.
\newblock
\urldef\tempurl%
\url{https://doi.org/10.1007/s10648-019-09514-z}
\showDOI{\tempurl}


\bibitem[\protect\citeauthoryear{Helsel}{Helsel}{1992}]%
        {vrandeducation}
\bibfield{author}{\bibinfo{person}{Sandra Helsel}.}
  \bibinfo{year}{1992}\natexlab{}.
\newblock \showarticletitle{Virtual Reality and Education}.
\newblock \bibinfo{journal}{\emph{Educational Technology}}
  \bibinfo{volume}{32}, \bibinfo{number}{5} (\bibinfo{year}{1992}),
  \bibinfo{pages}{38--42}.
\newblock
\showISSN{00131962}


\bibitem[\protect\citeauthoryear{Herrington, Reeves, and Oliver}{Herrington
  et~al\mbox{.}}{2007}]%
        {realism}
\bibfield{author}{\bibinfo{person}{Jan Herrington}, \bibinfo{person}{Thomas~C
  Reeves}, {and} \bibinfo{person}{Ron Oliver}.}
  \bibinfo{year}{2007}\natexlab{}.
\newblock \showarticletitle{Immersive learning technologies: Realism and online
  authentic learning}.
\newblock \bibinfo{journal}{\emph{Journal of Computing in Higher Education}}
  \bibinfo{volume}{19}, \bibinfo{number}{1} (\bibinfo{year}{2007}),
  \bibinfo{pages}{80--99}.
\newblock
\urldef\tempurl%
\url{https://doi.org/10.1007/BF03033421}
\showDOI{\tempurl}


\bibitem[\protect\citeauthoryear{Hirt, Eckard, and Kunz}{Hirt
  et~al\mbox{.}}{2020}]%
        {stress_vr}
\bibfield{author}{\bibinfo{person}{Christian Hirt}, \bibinfo{person}{Marcel
  Eckard}, {and} \bibinfo{person}{Andreas Kunz}.}
  \bibinfo{year}{2020}\natexlab{}.
\newblock \showarticletitle{Stress generation and non-intrusive measurement in
  virtual environments using eye tracking}.
\newblock \bibinfo{journal}{\emph{Journal of Ambient Intelligence and Humanized
  Computing}} \bibinfo{volume}{11}, \bibinfo{number}{1} (\bibinfo{year}{2020}),
  \bibinfo{pages}{1--13}.
\newblock
\urldef\tempurl%
\url{https://doi.org/10.1007/s12652-020-01845-y}
\showDOI{\tempurl}


\bibitem[\protect\citeauthoryear{Holmqvist, Nystr{\"o}m, Andersson, Dewhurst,
  Halszka, and {van de Weijer}}{Holmqvist et~al\mbox{.}}{2011}]%
        {holmqvist_book_eye_tracking}
\bibfield{author}{\bibinfo{person}{Kenneth Holmqvist}, \bibinfo{person}{Marcus
  Nystr{\"o}m}, \bibinfo{person}{Richard Andersson}, \bibinfo{person}{Richard
  Dewhurst}, \bibinfo{person}{Jarodzka Halszka}, {and} \bibinfo{person}{Joost
  {van de Weijer}}.} \bibinfo{year}{2011}\natexlab{}.
\newblock \bibinfo{booktitle}{\emph{Eye Tracking : A Comprehensive Guide to
  Methods and Measures}}.
\newblock \bibinfo{publisher}{Oxford University Press},
  \bibinfo{address}{United Kingdom}.
\newblock
\showISBNx{9780199697083}


\bibitem[\protect\citeauthoryear{Hone and {El Said}}{Hone and {El
  Said}}{2016}]%
        {mooc_social_interaction}
\bibfield{author}{\bibinfo{person}{Kate~S. Hone} {and}
  \bibinfo{person}{Ghada~R. {El Said}}.} \bibinfo{year}{2016}\natexlab{}.
\newblock \showarticletitle{Exploring the factors affecting MOOC retention: A
  survey study}.
\newblock \bibinfo{journal}{\emph{Computers \& Education}}
  \bibinfo{volume}{98} (\bibinfo{year}{2016}), \bibinfo{pages}{157--168}.
\newblock
\showISSN{0360-1315}
\urldef\tempurl%
\url{https://doi.org/10.1016/j.compedu.2016.03.016}
\showDOI{\tempurl}


\bibitem[\protect\citeauthoryear{Hu, Li, Zhang, Yi, Wang, and Manocha}{Hu
  et~al\mbox{.}}{2020}]%
        {8998375}
\bibfield{author}{\bibinfo{person}{Zhiming Hu}, \bibinfo{person}{Sheng Li},
  \bibinfo{person}{Congyi Zhang}, \bibinfo{person}{Kangrui Yi},
  \bibinfo{person}{Guoping Wang}, {and} \bibinfo{person}{Dinesh Manocha}.}
  \bibinfo{year}{2020}\natexlab{}.
\newblock \showarticletitle{DGaze: {CNN}-Based Gaze Prediction in Dynamic
  Scenes}.
\newblock \bibinfo{journal}{\emph{IEEE Transactions on Visualization and
  Computer Graphics}} \bibinfo{volume}{26}, \bibinfo{number}{5}
  (\bibinfo{year}{2020}), \bibinfo{pages}{1902--1911}.
\newblock
\urldef\tempurl%
\url{https://doi.org/10.1109/TVCG.2020.2973473}
\showDOI{\tempurl}


\bibitem[\protect\citeauthoryear{Jarodzka, Holmqvist, and Gruber}{Jarodzka
  et~al\mbox{.}}{2017}]%
        {Jarodzka_Holmqvist_Gruber_2017}
\bibfield{author}{\bibinfo{person}{Halszka Jarodzka}, \bibinfo{person}{Kenneth
  Holmqvist}, {and} \bibinfo{person}{Hans Gruber}.}
  \bibinfo{year}{2017}\natexlab{}.
\newblock \showarticletitle{Eye tracking in Educational Science: Theoretical
  frameworks and research agendas}.
\newblock \bibinfo{journal}{\emph{Journal of Eye Movement Research}}
  \bibinfo{volume}{10}, \bibinfo{number}{1} (\bibinfo{year}{2017}).
\newblock
\urldef\tempurl%
\url{https://doi.org/10.16910/jemr.10.1.3}
\showDOI{\tempurl}


\bibitem[\protect\citeauthoryear{Jo, Kim, and Kim}{Jo et~al\mbox{.}}{2016}]%
        {teleconference_twoaspects}
\bibfield{author}{\bibinfo{person}{Dongsik Jo}, \bibinfo{person}{Ki-Hong Kim},
  {and} \bibinfo{person}{Gerard~Jounghyun Kim}.}
  \bibinfo{year}{2016}\natexlab{}.
\newblock \showarticletitle{Effects of Avatar and Background Representation
  Forms to Co-Presence in Mixed Reality (MR) Tele-Conference Systems}. In
  \bibinfo{booktitle}{\emph{SIGGRAPH ASIA 2016 Virtual Reality Meets Physical
  Reality: Modelling and Simulating Virtual Humans and Environments}} (Macau).
  \bibinfo{publisher}{ACM}, \bibinfo{address}{New York, NY, USA}, Article
  \bibinfo{articleno}{12}, \bibinfo{numpages}{4}~pages.
\newblock
\showISBNx{9781450345484}
\urldef\tempurl%
\url{https://doi.org/10.1145/2992138.2992146}
\showDOI{\tempurl}


\bibitem[\protect\citeauthoryear{Just and Carpenter}{Just and
  Carpenter}{1976}]%
        {just1976eye}
\bibfield{author}{\bibinfo{person}{Marcel~A. Just} {and}
  \bibinfo{person}{Patricia~A. Carpenter}.} \bibinfo{year}{1976}\natexlab{}.
\newblock \showarticletitle{Eye fixations and cognitive processes}.
\newblock \bibinfo{journal}{\emph{Cognitive Psychology}} \bibinfo{volume}{8},
  \bibinfo{number}{4} (\bibinfo{year}{1976}), \bibinfo{pages}{441--480}.
\newblock
\showISSN{0010-0285}
\urldef\tempurl%
\url{https://doi.org/10.1016/0010-0285(76)90015-3}
\showDOI{\tempurl}


\bibitem[\protect\citeauthoryear{Kantonen, Woodward, and Katz}{Kantonen
  et~al\mbox{.}}{2010}]%
        {teleconference_VR}
\bibfield{author}{\bibinfo{person}{Tuomas Kantonen}, \bibinfo{person}{Charles
  Woodward}, {and} \bibinfo{person}{Neil Katz}.}
  \bibinfo{year}{2010}\natexlab{}.
\newblock \showarticletitle{Mixed reality in virtual world teleconferencing}.
  In \bibinfo{booktitle}{\emph{2010 IEEE Virtual Reality Conference (VR)}}
  (Waltham, MA, USA). \bibinfo{publisher}{IEEE}, \bibinfo{address}{New York,
  NY, USA}, \bibinfo{pages}{179--182}.
\newblock
\urldef\tempurl%
\url{https://doi.org/10.1109/VR.2010.5444792}
\showDOI{\tempurl}


\bibitem[\protect\citeauthoryear{Khan, Li, and R{\'e}hman}{Khan
  et~al\mbox{.}}{2016}]%
        {teleconference_immersion}
\bibfield{author}{\bibinfo{person}{Muhammad Sikandar~Lal Khan},
  \bibinfo{person}{Haibo Li}, {and} \bibinfo{person}{Shafiq~Ur R{\'e}hman}.}
  \bibinfo{year}{2016}\natexlab{}.
\newblock \showarticletitle{Tele-Immersion: Virtual Reality Based
  Collaboration}. In \bibinfo{booktitle}{\emph{International Conference on
  Human-Computer Interaction}} (Toronto, Canada).
  \bibinfo{publisher}{Springer}, \bibinfo{address}{Cham},
  \bibinfo{pages}{352--357}.
\newblock
\urldef\tempurl%
\url{https://doi.org/10.1007/978-3-319-40548-3_59}
\showDOI{\tempurl}


\bibitem[\protect\citeauthoryear{Lamb and Etopio}{Lamb and Etopio}{2020}]%
        {lambvirtual}
\bibfield{author}{\bibinfo{person}{Richard Lamb} {and}
  \bibinfo{person}{Elisabeth~A. Etopio}.} \bibinfo{year}{2020}\natexlab{}.
\newblock \showarticletitle{{Virtual Reality: A Tool for Preservice Science
  Teachers to Put Theory into Practice}}.
\newblock \bibinfo{journal}{\emph{Journal of Science Education and Technology}}
  \bibinfo{volume}{29}, \bibinfo{number}{4} (\bibinfo{year}{2020}),
  \bibinfo{pages}{573--585}.
\newblock
\urldef\tempurl%
\url{https://doi.org/10.1007/s10956-020-09837-5}
\showDOI{\tempurl}


\bibitem[\protect\citeauthoryear{Lan, Luo, and Hao}{Lan et~al\mbox{.}}{2016}]%
        {teleconference_realtime}
\bibfield{author}{\bibinfo{person}{Gongjin Lan}, \bibinfo{person}{Ziyun Luo},
  {and} \bibinfo{person}{Qi Hao}.} \bibinfo{year}{2016}\natexlab{}.
\newblock \showarticletitle{Development of a virtual reality teleconference
  system using distributed depth sensors}. In \bibinfo{booktitle}{\emph{2016
  2nd IEEE International Conference on Computer and Communications (ICCC)}}
  (Chengdu, China). \bibinfo{publisher}{IEEE}, \bibinfo{address}{New York, NY,
  USA}, \bibinfo{pages}{975--978}.
\newblock
\urldef\tempurl%
\url{https://doi.org/10.1109/CompComm.2016.7924850}
\showDOI{\tempurl}


\bibitem[\protect\citeauthoryear{Lang, Wei, Xu, Zhao, and Yu}{Lang
  et~al\mbox{.}}{2018}]%
        {8448290}
\bibfield{author}{\bibinfo{person}{Yining Lang}, \bibinfo{person}{Liang Wei},
  \bibinfo{person}{Fang Xu}, \bibinfo{person}{Yibiao Zhao}, {and}
  \bibinfo{person}{Lap-Fai Yu}.} \bibinfo{year}{2018}\natexlab{}.
\newblock \showarticletitle{Synthesizing Personalized Training Programs for
  Improving Driving Habits via Virtual Reality}. In
  \bibinfo{booktitle}{\emph{2018 IEEE Conference on Virtual Reality and 3D User
  Interfaces (VR)}} (Reutlingen, Germany). \bibinfo{publisher}{IEEE},
  \bibinfo{address}{New York, NY, USA}, \bibinfo{pages}{297--304}.
\newblock
\urldef\tempurl%
\url{https://doi.org/10.1109/VR.2018.8448290}
\showDOI{\tempurl}


\bibitem[\protect\citeauthoryear{Liao, Sung, Wang, and Lin}{Liao
  et~al\mbox{.}}{2019}]%
        {8797708}
\bibfield{author}{\bibinfo{person}{Meng-Yun Liao}, \bibinfo{person}{Ching-Ying
  Sung}, \bibinfo{person}{Hao-Chuan Wang}, {and} \bibinfo{person}{Wen-Chieh
  Lin}.} \bibinfo{year}{2019}\natexlab{}.
\newblock \showarticletitle{Virtual Classmates: Embodying Historical Learners'
  Messages as Learning Companions in a VR Classroom through Comment Mapping}.
  In \bibinfo{booktitle}{\emph{2019 IEEE Conference on Virtual Reality and 3D
  User Interfaces (VR)}} (Osaka, Japan). \bibinfo{publisher}{IEEE},
  \bibinfo{address}{New York, NY, USA}, \bibinfo{pages}{163--171}.
\newblock
\urldef\tempurl%
\url{https://doi.org/10.1109/VR.2019.8797708}
\showDOI{\tempurl}


\bibitem[\protect\citeauthoryear{Lombard, Bolmarcich, and Weinstein}{Lombard
  et~al\mbox{.}}{2009}]%
        {lombard_presence}
\bibfield{author}{\bibinfo{person}{Matthew Lombard}, \bibinfo{person}{Theresa
  Bolmarcich}, {and} \bibinfo{person}{Lisa Weinstein}.}
  \bibinfo{year}{2009}\natexlab{}.
\newblock \showarticletitle{Measuring Presence: The Temple Presence Inventory}.
  In \bibinfo{booktitle}{\emph{Proceedings of the 12th Annual International
  Workshop on Presence}}. \bibinfo{publisher}{The International Society for
  Presence Research}, \bibinfo{address}{Los Angeles, CA, USA},
  \bibinfo{pages}{1--15}.
\newblock


\bibitem[\protect\citeauthoryear{Mangalmurti, Kistler, Quarrie, Sharp, Persky,
  and Shaw}{Mangalmurti et~al\mbox{.}}{2020}]%
        {mangalmurti_2020}
\bibfield{author}{\bibinfo{person}{Aman Mangalmurti}, \bibinfo{person}{William
  Kistler}, \bibinfo{person}{Barrington Quarrie}, \bibinfo{person}{Wendy
  Sharp}, \bibinfo{person}{Susan Persky}, {and} \bibinfo{person}{Philip Shaw}.}
  \bibinfo{year}{2020}\natexlab{}.
\newblock \showarticletitle{Using virtual reality to define the mechanisms
  linking symptoms with cognitive deficits in attention deficit hyperactivity
  disorder}.
\newblock \bibinfo{journal}{\emph{Scientific Reports}}  \bibinfo{volume}{10}
  (\bibinfo{date}{12} \bibinfo{year}{2020}).
\newblock
\urldef\tempurl%
\url{https://doi.org/10.1038/s41598-019-56936-4}
\showDOI{\tempurl}


\bibitem[\protect\citeauthoryear{Marks, Sibley, and {J. Ben Arbaugh}}{Marks
  et~al\mbox{.}}{2005}]%
        {doi:10.1177/1052562904271199}
\bibfield{author}{\bibinfo{person}{Ronald~B. Marks},
  \bibinfo{person}{Stanley~D. Sibley}, {and} \bibinfo{person}{{J. Ben
  Arbaugh}}.} \bibinfo{year}{2005}\natexlab{}.
\newblock \showarticletitle{A Structural Equation Model of Predictors for
  Effective Online Learning}.
\newblock \bibinfo{journal}{\emph{Journal of Management Education}}
  \bibinfo{volume}{29}, \bibinfo{number}{4} (\bibinfo{year}{2005}),
  \bibinfo{pages}{531--563}.
\newblock
\urldef\tempurl%
\url{https://doi.org/10.1177/1052562904271199}
\showDOI{\tempurl}


\bibitem[\protect\citeauthoryear{Math{\^o}t, Fabius, Van~Heusden, and Van~der
  Stigchel}{Math{\^o}t et~al\mbox{.}}{2018}]%
        {Mathot2018}
\bibfield{author}{\bibinfo{person}{Sebastiaan Math{\^o}t},
  \bibinfo{person}{Jasper Fabius}, \bibinfo{person}{Elle Van~Heusden}, {and}
  \bibinfo{person}{Stefan Van~der Stigchel}.} \bibinfo{year}{2018}\natexlab{}.
\newblock \showarticletitle{Safe and sensible preprocessing and baseline
  correction of pupil-size data}.
\newblock \bibinfo{journal}{\emph{Behavior Research Methods}}
  \bibinfo{volume}{50}, \bibinfo{number}{1} (\bibinfo{year}{2018}),
  \bibinfo{pages}{94--106}.
\newblock
\showISSN{1554-3528}
\urldef\tempurl%
\url{https://doi.org/10.3758/s13428-017-1007-2}
\showDOI{\tempurl}


\bibitem[\protect\citeauthoryear{Mayer}{Mayer}{2010}]%
        {MAYER2010167}
\bibfield{author}{\bibinfo{person}{Richard~E. Mayer}.}
  \bibinfo{year}{2010}\natexlab{}.
\newblock \showarticletitle{Unique contributions of eye-tracking research to
  the study of learning with graphics}.
\newblock \bibinfo{journal}{\emph{Learning and Instruction}}
  \bibinfo{volume}{20}, \bibinfo{number}{2} (\bibinfo{year}{2010}),
  \bibinfo{pages}{167--171}.
\newblock
\showISSN{0959-4752}
\urldef\tempurl%
\url{https://doi.org/10.1016/j.learninstruc.2009.02.012}
\showDOI{\tempurl}


\bibitem[\protect\citeauthoryear{Moro, \v{S}tromberga, Raikos, and
  Stirling}{Moro et~al\mbox{.}}{2017}]%
        {doi:10.1002/ase.1696}
\bibfield{author}{\bibinfo{person}{Christian Moro}, \bibinfo{person}{Zane
  \v{S}tromberga}, \bibinfo{person}{Athanasios Raikos}, {and}
  \bibinfo{person}{Allan Stirling}.} \bibinfo{year}{2017}\natexlab{}.
\newblock \showarticletitle{The effectiveness of virtual and augmented reality
  in health sciences and medical anatomy}.
\newblock \bibinfo{journal}{\emph{Anatomical Sciences Education}}
  \bibinfo{volume}{10}, \bibinfo{number}{6} (\bibinfo{year}{2017}),
  \bibinfo{pages}{549--559}.
\newblock
\urldef\tempurl%
\url{https://doi.org/10.1002/ase.1696}
\showDOI{\tempurl}


\bibitem[\protect\citeauthoryear{Negi and Mitra}{Negi and Mitra}{2020}]%
        {Negi_Mitra_2020}
\bibfield{author}{\bibinfo{person}{Shivsevak Negi} {and}
  \bibinfo{person}{Ritayan Mitra}.} \bibinfo{year}{2020}\natexlab{}.
\newblock \showarticletitle{Fixation duration and the learning process: an eye
  tracking study with subtitled videos}.
\newblock \bibinfo{journal}{\emph{Journal of Eye Movement Research}}
  \bibinfo{volume}{13}, \bibinfo{number}{6} (\bibinfo{year}{2020}).
\newblock
\urldef\tempurl%
\url{https://doi.org/10.16910/jemr.13.6.1}
\showDOI{\tempurl}


\bibitem[\protect\citeauthoryear{Nolin, Stipanicic, Henry, Lachapelle,
  Lussier-Desrochers, Rizzo, and Allain}{Nolin et~al\mbox{.}}{2016}]%
        {Nolin2016ClinicaVRCA}
\bibfield{author}{\bibinfo{person}{Pierre Nolin}, \bibinfo{person}{Annie
  Stipanicic}, \bibinfo{person}{Myl{\`e}ne Henry}, \bibinfo{person}{Yves
  Lachapelle}, \bibinfo{person}{Dany Lussier-Desrochers},
  \bibinfo{person}{Albert~S. Rizzo}, {and} \bibinfo{person}{Philippe Allain}.}
  \bibinfo{year}{2016}\natexlab{}.
\newblock \showarticletitle{Clinica{VR}: Classroom-{CPT}: A virtual reality
  tool for assessing attention and inhibition in children and adolescents}.
\newblock \bibinfo{journal}{\emph{Computers in Human Behavior}}
  \bibinfo{volume}{59} (\bibinfo{year}{2016}), \bibinfo{pages}{327--333}.
\newblock
\urldef\tempurl%
\url{https://doi.org/10.1016/j.chb.2016.02.023}
\showDOI{\tempurl}


\bibitem[\protect\citeauthoryear{Olmos-Raya, Ferreira-Cavalcanti, Contero,
  Castellanos, Giglioli, and Alca{\~n}iz}{Olmos-Raya et~al\mbox{.}}{2018}]%
        {motivation_immersion}
\bibfield{author}{\bibinfo{person}{Elena Olmos-Raya}, \bibinfo{person}{Janaina
  Ferreira-Cavalcanti}, \bibinfo{person}{Manuel Contero},
  \bibinfo{person}{M~Concepci{\'o}n Castellanos}, \bibinfo{person}{Irene
  Alice~Chicchi Giglioli}, {and} \bibinfo{person}{Mariano Alca{\~n}iz}.}
  \bibinfo{year}{2018}\natexlab{}.
\newblock \showarticletitle{Mobile virtual reality as an educational platform:
  A pilot study on the impact of immersion and positive emotion induction in
  the learning process}.
\newblock \bibinfo{journal}{\emph{EURASIA Journal of Mathematics, Science and
  Technology Education}} \bibinfo{volume}{14}, \bibinfo{number}{6}
  (\bibinfo{year}{2018}), \bibinfo{pages}{2045--2057}.
\newblock
\urldef\tempurl%
\url{https://doi.org/10.29333/ejmste/85874}
\showDOI{\tempurl}


\bibitem[\protect\citeauthoryear{Orlosky, Itoh, Ranchet, Kiyokawa, Morgan, and
  Devos}{Orlosky et~al\mbox{.}}{2017}]%
        {7829437}
\bibfield{author}{\bibinfo{person}{Jason Orlosky}, \bibinfo{person}{Yuta Itoh},
  \bibinfo{person}{Maud Ranchet}, \bibinfo{person}{Kiyoshi Kiyokawa},
  \bibinfo{person}{John Morgan}, {and} \bibinfo{person}{Hannes Devos}.}
  \bibinfo{year}{2017}\natexlab{}.
\newblock \showarticletitle{Emulation of Physician Tasks in Eye-Tracked Virtual
  Reality for Remote Diagnosis of Neurodegenerative Disease}.
\newblock \bibinfo{journal}{\emph{IEEE Transactions on Visualization and
  Computer Graphics}} \bibinfo{volume}{23}, \bibinfo{number}{4}
  (\bibinfo{year}{2017}), \bibinfo{pages}{1302--1311}.
\newblock
\urldef\tempurl%
\url{https://doi.org/10.1109/TVCG.2017.2657018}
\showDOI{\tempurl}


\bibitem[\protect\citeauthoryear{Pomplun, Garaas, and Carrasco}{Pomplun
  et~al\mbox{.}}{2013}]%
        {task_difficulty_fixation_durations}
\bibfield{author}{\bibinfo{person}{Marc Pomplun}, \bibinfo{person}{Tyler
  Garaas}, {and} \bibinfo{person}{Marisa Carrasco}.}
  \bibinfo{year}{2013}\natexlab{}.
\newblock \showarticletitle{The effects of task difficulty on visual search
  strategy in virtual 3D displays}.
\newblock \bibinfo{journal}{\emph{Journal of vision}}  \bibinfo{volume}{13}
  (\bibinfo{year}{2013}).
\newblock
\urldef\tempurl%
\url{https://doi.org/10.1167/13.3.24}
\showDOI{\tempurl}


\bibitem[\protect\citeauthoryear{Psotka}{Psotka}{1995}]%
        {vreducationdomain}
\bibfield{author}{\bibinfo{person}{Joseph Psotka}.}
  \bibinfo{year}{1995}\natexlab{}.
\newblock \showarticletitle{Immersive training systems: Virtual reality and
  education and training}.
\newblock \bibinfo{journal}{\emph{Instructional science}} \bibinfo{volume}{23},
  \bibinfo{number}{5-6} (\bibinfo{year}{1995}), \bibinfo{pages}{405--431}.
\newblock
\urldef\tempurl%
\url{https://doi.org/10.1007/BF00896880}
\showDOI{\tempurl}


\bibitem[\protect\citeauthoryear{Rappa, Ledger, Teo, Wong, Power, and
  Hilliard}{Rappa et~al\mbox{.}}{2019}]%
        {vr_eyetracking_review}
\bibfield{author}{\bibinfo{person}{Natasha~Anne Rappa}, \bibinfo{person}{Susan
  Ledger}, \bibinfo{person}{Timothy Teo}, \bibinfo{person}{Kok~Wai Wong},
  \bibinfo{person}{Brad Power}, {and} \bibinfo{person}{Bruce Hilliard}.}
  \bibinfo{year}{2019}\natexlab{}.
\newblock \showarticletitle{The use of eye tracking technology to explore
  learning and performance within virtual reality and mixed reality settings: a
  scoping review}.
\newblock \bibinfo{journal}{\emph{Interactive Learning Environments}}
  \bibinfo{volume}{0}, \bibinfo{number}{0} (\bibinfo{year}{2019}),
  \bibinfo{pages}{1--13}.
\newblock
\urldef\tempurl%
\url{https://doi.org/10.1080/10494820.2019.1702560}
\showDOI{\tempurl}


\bibitem[\protect\citeauthoryear{Ray and Deb}{Ray and Deb}{2016}]%
        {smartphone_vr}
\bibfield{author}{\bibinfo{person}{Ananda~Bibek Ray} {and}
  \bibinfo{person}{Suman Deb}.} \bibinfo{year}{2016}\natexlab{}.
\newblock \showarticletitle{Smartphone Based Virtual Reality Systems in
  Classroom Teaching — A Study on the Effects of Learning Outcome}. In
  \bibinfo{booktitle}{\emph{2016 IEEE Eighth International Conference on
  Technology for Education (T4E)}} (Mumbai, India). \bibinfo{publisher}{IEEE},
  \bibinfo{address}{New York, NY, USA}, \bibinfo{pages}{68--71}.
\newblock
\urldef\tempurl%
\url{https://doi.org/10.1109/T4E.2016.022}
\showDOI{\tempurl}


\bibitem[\protect\citeauthoryear{Rizzo, Bowerly, Buckwalter, Klimchuk, Mitura,
  and Parsons}{Rizzo et~al\mbox{.}}{2006}]%
        {rizzo_bowerly_buckwalter_klimchuk_mitura_parsons_2009}
\bibfield{author}{\bibinfo{person}{Albert~A. Rizzo}, \bibinfo{person}{Todd
  Bowerly}, \bibinfo{person}{J.~Galen Buckwalter}, \bibinfo{person}{Dean
  Klimchuk}, \bibinfo{person}{Roman Mitura}, {and} \bibinfo{person}{Thomas~D.
  Parsons}.} \bibinfo{year}{2006}\natexlab{}.
\newblock \showarticletitle{A Virtual Reality Scenario for All Seasons: The
  Virtual Classroom}.
\newblock \bibinfo{journal}{\emph{CNS Spectrums}} \bibinfo{volume}{11},
  \bibinfo{number}{1} (\bibinfo{year}{2006}), \bibinfo{pages}{35–44}.
\newblock
\urldef\tempurl%
\url{https://doi.org/10.1017/S1092852900024196}
\showDOI{\tempurl}


\bibitem[\protect\citeauthoryear{Salvucci and Goldberg}{Salvucci and
  Goldberg}{2000}]%
        {salvucci2000identifying}
\bibfield{author}{\bibinfo{person}{Dario~D. Salvucci} {and}
  \bibinfo{person}{Joseph~H. Goldberg}.} \bibinfo{year}{2000}\natexlab{}.
\newblock \showarticletitle{Identifying Fixations and Saccades in Eye-Tracking
  Protocols}. In \bibinfo{booktitle}{\emph{Proceedings of the 2000 Symposium on
  Eye Tracking Research \& Applications}} (Palm Beach Gardens, FL, USA).
  \bibinfo{publisher}{ACM}, \bibinfo{address}{New York, NY, USA},
  \bibinfo{pages}{71–78}.
\newblock
\showISBNx{1581132808}
\urldef\tempurl%
\url{https://doi.org/10.1145/355017.355028}
\showDOI{\tempurl}


\bibitem[\protect\citeauthoryear{Savitzky and Golay}{Savitzky and
  Golay}{1964}]%
        {savitzky64}
\bibfield{author}{\bibinfo{person}{Abraham Savitzky} {and}
  \bibinfo{person}{Marcel J.~E. Golay}.} \bibinfo{year}{1964}\natexlab{}.
\newblock \showarticletitle{Smoothing and Differentiation of Data by Simplified
  Least Squares Procedures}.
\newblock \bibinfo{journal}{\emph{Analytical Chemistry}}  \bibinfo{volume}{36}
  (\bibinfo{year}{1964}), \bibinfo{pages}{1627--1639}.
\newblock
\urldef\tempurl%
\url{https://doi.org/10.1021/ac60214a047}
\showDOI{\tempurl}


\bibitem[\protect\citeauthoryear{Schubert, Friedmann, and Regenbrecht}{Schubert
  et~al\mbox{.}}{2001}]%
        {schubert_presence}
\bibfield{author}{\bibinfo{person}{Thomas Schubert}, \bibinfo{person}{Frank
  Friedmann}, {and} \bibinfo{person}{Holger Regenbrecht}.}
  \bibinfo{year}{2001}\natexlab{}.
\newblock \showarticletitle{The Experience of Presence: Factor Analytic
  Insights}.
\newblock \bibinfo{journal}{\emph{Presence}} \bibinfo{volume}{10},
  \bibinfo{number}{3} (\bibinfo{year}{2001}), \bibinfo{pages}{266--281}.
\newblock
\urldef\tempurl%
\url{https://doi.org/10.1162/105474601300343603}
\showDOI{\tempurl}


\bibitem[\protect\citeauthoryear{Seo, Kim, Mundy, Heo, and Kim}{Seo
  et~al\mbox{.}}{2019}]%
        {Seo2019JointAV}
\bibfield{author}{\bibinfo{person}{Seung-hun Seo}, \bibinfo{person}{Eunjoo
  Kim}, \bibinfo{person}{Peter Mundy}, \bibinfo{person}{Jiwoong Heo}, {and}
  \bibinfo{person}{Kwanguk Kim}.} \bibinfo{year}{2019}\natexlab{}.
\newblock \showarticletitle{Joint Attention Virtual Classroom: A Preliminary
  Study}.
\newblock \bibinfo{journal}{\emph{Psychiatry Investigation}}
  \bibinfo{volume}{16} (\bibinfo{year}{2019}), \bibinfo{pages}{292--299}.
\newblock
\urldef\tempurl%
\url{https://doi.org/10.30773/pi.2019.02.08}
\showDOI{\tempurl}


\bibitem[\protect\citeauthoryear{Sharma, Agada, and Ruffin}{Sharma
  et~al\mbox{.}}{2013}]%
        {vrclassroomconstructivist}
\bibfield{author}{\bibinfo{person}{Sharad Sharma}, \bibinfo{person}{Ruth
  Agada}, {and} \bibinfo{person}{Jeff Ruffin}.}
  \bibinfo{year}{2013}\natexlab{}.
\newblock \showarticletitle{Virtual reality classroom as an constructivist
  approach}. In \bibinfo{booktitle}{\emph{2013 Proceedings of IEEE
  Southeastcon}} (Jacksonville, FL, USA). \bibinfo{publisher}{IEEE},
  \bibinfo{address}{New York, NY, USA}, \bibinfo{pages}{1--5}.
\newblock
\showISBNx{978-1-4799-0052-7}
\urldef\tempurl%
\url{https://doi.org/10.1109/SECON.2013.6567441}
\showDOI{\tempurl}


\bibitem[\protect\citeauthoryear{Shavelson, Hubner, and Stanton}{Shavelson
  et~al\mbox{.}}{1976}]%
        {self_concept_1976}
\bibfield{author}{\bibinfo{person}{Richard~J. Shavelson},
  \bibinfo{person}{Judith~J. Hubner}, {and} \bibinfo{person}{George~C.
  Stanton}.} \bibinfo{year}{1976}\natexlab{}.
\newblock \showarticletitle{Self-Concept: Validation of Construct
  Interpretations}.
\newblock \bibinfo{journal}{\emph{Review of Educational Research}}
  \bibinfo{volume}{46}, \bibinfo{number}{3} (\bibinfo{year}{1976}),
  \bibinfo{pages}{407--441}.
\newblock
\urldef\tempurl%
\url{https://doi.org/10.3102/00346543046003407}
\showDOI{\tempurl}


\bibitem[\protect\citeauthoryear{Simeone, Speicher, Molnar, Wilde, and
  Daiber}{Simeone et~al\mbox{.}}{2019}]%
        {livehumanrole}
\bibfield{author}{\bibinfo{person}{Adalberto~L. Simeone},
  \bibinfo{person}{Marco Speicher}, \bibinfo{person}{Andreea Molnar},
  \bibinfo{person}{Adriana Wilde}, {and} \bibinfo{person}{Florian Daiber}.}
  \bibinfo{year}{2019}\natexlab{}.
\newblock \showarticletitle{LIVE: The Human Role in Learning in Immersive
  Virtual Environments}. In \bibinfo{booktitle}{\emph{Symposium on Spatial User
  Interaction}} (New Orleans, LA, USA). \bibinfo{publisher}{ACM},
  \bibinfo{address}{New York, NY, USA}, Article \bibinfo{articleno}{5},
  \bibinfo{numpages}{11}~pages.
\newblock
\showISBNx{9781450369756}
\urldef\tempurl%
\url{https://doi.org/10.1145/3357251.3357590}
\showDOI{\tempurl}


\bibitem[\protect\citeauthoryear{S{\"u}mer, Goldberg, St{\"u}rmer, Seidel,
  Gerjets, Trautwein, and Kasneci}{S{\"u}mer et~al\mbox{.}}{2018}]%
        {Sumer_2018_CVPR_Workshops}
\bibfield{author}{\bibinfo{person}{{\"O}mer S{\"u}mer},
  \bibinfo{person}{Patricia Goldberg}, \bibinfo{person}{Kathleen St{\"u}rmer},
  \bibinfo{person}{Tina Seidel}, \bibinfo{person}{Peter Gerjets},
  \bibinfo{person}{Ulrich Trautwein}, {and} \bibinfo{person}{Enkelejda
  Kasneci}.} \bibinfo{year}{2018}\natexlab{}.
\newblock \showarticletitle{Teachers' Perception in the Classroom}. In
  \bibinfo{booktitle}{\emph{Proceedings of the IEEE Conference on Computer
  Vision and Pattern Recognition (CVPR) Workshops}} (Salt Lake City, UT, USA).
  \bibinfo{publisher}{IEEE}, \bibinfo{address}{New York, NY, USA},
  \bibinfo{pages}{2315--2324}.
\newblock


\bibitem[\protect\citeauthoryear{Weintrop, Beheshti, Horn, Kai, Jona, Trouille,
  and Wilensky}{Weintrop et~al\mbox{.}}{2016}]%
        {Weintrop2016DefiningCT}
\bibfield{author}{\bibinfo{person}{David Weintrop}, \bibinfo{person}{Elham
  Beheshti}, \bibinfo{person}{Michael Horn}, \bibinfo{person}{Orton Kai},
  \bibinfo{person}{Kemi Jona}, \bibinfo{person}{Laura Trouille}, {and}
  \bibinfo{person}{Uri Wilensky}.} \bibinfo{year}{2016}\natexlab{}.
\newblock \showarticletitle{Defining Computational Thinking for Mathematics and
  Science Classrooms}.
\newblock \bibinfo{journal}{\emph{Journal of Science Education and Technology}}
   \bibinfo{volume}{25} (\bibinfo{year}{2016}), \bibinfo{pages}{127--147}.
\newblock
\urldef\tempurl%
\url{https://doi.org/10.1007/s10956-015-9581-5}
\showDOI{\tempurl}


\bibitem[\protect\citeauthoryear{Wobbrock, Findlater, Gergle, and
  Higgins}{Wobbrock et~al\mbox{.}}{2011}]%
        {10.1145/1978942.1978963}
\bibfield{author}{\bibinfo{person}{Jacob~O. Wobbrock}, \bibinfo{person}{Leah
  Findlater}, \bibinfo{person}{Darren Gergle}, {and} \bibinfo{person}{James~J.
  Higgins}.} \bibinfo{year}{2011}\natexlab{}.
\newblock \showarticletitle{The Aligned Rank Transform for Nonparametric
  Factorial Analyses Using Only Anova Procedures}. In
  \bibinfo{booktitle}{\emph{Proceedings of the SIGCHI Conference on Human
  Factors in Computing Systems}} (Vancouver, Canada). \bibinfo{publisher}{ACM},
  \bibinfo{address}{New York, NY, USA}, \bibinfo{pages}{143–146}.
\newblock
\showISBNx{9781450302289}
\urldef\tempurl%
\url{https://doi.org/10.1145/1978942.1978963}
\showDOI{\tempurl}


\bibitem[\protect\citeauthoryear{Youngblut}{Youngblut}{1998}]%
        {youngblut1998educational}
\bibfield{author}{\bibinfo{person}{Christine Youngblut}.}
  \bibinfo{year}{1998}\natexlab{}.
\newblock \bibinfo{booktitle}{\emph{Educational uses of virtual reality
  technology}}.
\newblock \bibinfo{type}{{T}echnical {R}eport}. \bibinfo{institution}{Institute
  for Defense Analyses}, \bibinfo{address}{Alexandria, VA, USA}.
\newblock


\end{thebibliography}

\end{document}